\documentclass[%
 aip,
 jmp,%
 amsmath,amssymb,
 preprint,%
]{revtex4-1}

\usepackage{xcolor}
\usepackage{graphicx}
\usepackage{dcolumn}
\usepackage{bm}

\def \a {\alpha}
\def \b {\beta}
\def \g {\gamma}
\def \d {\delta}

\def \f {\varphi}
\def \k {\kappa}
\def \l {\lambda}
\def \m {\mu}
\def \n {\nu}
\def \r {\rho}
\def \s {\sigma}
\def \t {\tau}
\def \x {\xi}

\def \G {\Gamma}
\def \D {\Delta}

\def \pd {\partial}
\def \L {\hat{\mathcal{L}}}
\def \Lh {\mathcal{L}}
\def \half {\frac{1}{2}}

\def \bn {\bar{\nabla}}
\def \nb {\nabla}

\newcommand*{\citen}{}
\DeclareRobustCommand*{\citen}[1]{%
  \begingroup
    \romannumeral-`\x 
    \setcitestyle{numbers}%
    \cite{#1}%
  \endgroup
}%

\begin{document}


\title[Covariant conserved currents for scalar-tensor Horndeski theory]{Covariant conserved currents for scalar-tensor Horndeski theory}

\author{J. Schmidt}
\email{schmijos@fjfi.cvut.cz}
\affiliation{ 
Department of Physics, Faculty of Nuclear Physics and Physical Engineering, Czech Technical University in Prague, B\v{r}ehov\'{a} 7, 115 19 Praha 1, Czech Republic
}%

\author{J. Bi\v{c}\'{a}k}
\email{Jiri.Bicak@mff.cuni.cz}
\affiliation{%
 Institute of Theoretical Physics, Faculty of Mathematics and Physics, Charles University, V Hole\v{s}ovi\v{c}k\'{a}ch 2, 180 00 Praha 8, Czech Republic
}%
\altaffiliation[Also at ]{Max Planck Institute for Gravitational Physics, Albert Einstein Institute, Am M\"{u}hlenberg 1, D-14476 Golm, Germany}


\begin{abstract}
The scalar-tensor theories have become popular recently in particular in connection with attempts to explain present accelerated expansion of the universe, but they have been considered as a natural extension of general relativity long time ago. The Horndeski scalar-tensor theory involving four invariantly defined Lagrangians is a natural choice since it implies field equations involving at most second derivatives. Following the formalisms of defining covariant global quantities and conservation laws for perturbations of spacetimes in standard general relativity we extend these methods to the general Horndeski theory and find the covariant conserved currents for all four Lagrangians. The current is also constructed in the case of linear perturbations involving both metric and scalar field. As a specific illustration we derive a superpotential which leads to the covariantly conserved current in the Branse-Dicke theory.
%
\end{abstract}

\maketitle

\section{Introduction}

The modifications and generalizations of Einstein's theory of gravitation have been studied very actively in recent decades primarily in cosmological contexts with an attempt to explain the present accelerated expansions of the universe (for reviews, see, e.g., Refs. \citen{pap}-\citen{sotiriou}). The motivation for such studies is also coming from observations on galactic scales related to galaxy rotation curves and the corresponding problem of dark matter. Most recently, in addition, the detection of gravitational waves from black hole mergers gives prospects to investigate the possible deviations from Einstein's theory in local strong-gravity regimes. 

The modification of Einstein's theory by introducing a scalar field together with metric for the description of gravity appears to be most natural and most widely discussed. 

The purpose of the present work is to develop the formalism for derivation of covariant conserved currents in the general case of the scalar-tensor Horndeski theory.\cite{horndeski} Our formalism is inspired by the original method of derivation of covariant currents in the classical general relativity by Katz, Bi\v{c}\'{a}k, and Lynden-Bell (KBL).\cite{KBL} This method of formulation of conservation laws with respect to (w.r.t.) general curved backgrounds was developed in order to understand how Mach's principle can be formulated within cosmological perturbation theory. The resulting superpotential, from which the conserved quantities for arbitrarily \textit{large} perturbations can be expressed as surface integrals, was found after applying certain natural criteria. It is commonly called the ``KBL superpotential'' as first designated by Julia and Silva.\cite{JuSi1}$^,$\cite{JuSi2} It is unambiguous and satisfactory in spacetimes with or without a cosmological constant in any spacetime dimension $D \geq 3$ (see Refs. \citen{JuSi1} and \citen{JuSi2}). It was applied in a number of situations, for example, in the studies of the causal generation of cosmological perturbations determining large-scale structure formation and in the problem of the backreaction in slow-roll inflation; see Refs. \citen{AmShe} and \citen{LoUn} and references therein.

In the KBL formalism a strong conservation law in spacetime $M$ with metric $g_{\m\n}$ is derived with respect to a background spacetime $\bar{M}$ with metric $\bar{g}_{\m\n}$; for details, see Refs. \citen{KBL} and \citen{PeKa}. The background (auxiliary) metric enables one to define covariantly global quantities at infinity (asymptotically flat, de Sitter, anti de Sitter, etc.). The background metric is then to be considered only at the ``boundary.'' When perturbations of a given spacetime are studied, the background metric plays a role in the ``interior'' as well. 

Before we turn to the characterization of the Horndeski theory and the construction of the conserved currents, we briefly recall the KBL formalism. The spacetimes $M$ and $\bar{M}$ are connected via diffeomorphism $\phi$; once this is fixed, we can consider both metrics to be on $M$ (the pullback $\phi^* \bar{g}_{\m\n}$ is denoted just by $\bar{g}_{\m\n}$). Correspondingly, the covariant derivatives, the Christoffel symbols, and the curvature tensors are denoted by $\nb_{\a}$, $\G^{\l}_{\m\n}$, $R{^{\l}}_{\t\r\s}$ and by $\bn_{\a}$, $\bar{\G}^{\l}_{\m\n}$, $\bar{R}{^{\l}}_{\t\r\s}$, the differences of the Christoffel symbols will be denoted by $\D^{\l}_{\m\n} = \G^{\l}_{\m\n} - \bar{\G}^{\l}_{\m\n}$. Following KBL, the Lagrangian density for (not necessarily small) perturbations is written as
\begin{equation}
	\L_G = \L - \bar{\L}, \quad \L = - \frac{1}{2\k} \left( \hat{R} + \pd_{\m} \hat{k}^{\m} \right), \quad \bar{\L} = - \frac{1}{2\k} \bar{\hat{R}}, \label{KBLLg}
\end{equation}
where $\hat{R} = \sqrt{-g} R$ is the scalar density curvature of spacetime $M$, $\bar{\hat{R}} = \sqrt{-\bar{g}} \bar{R}$ that of $\bar{M}$, $\hat{k}^{\m}$ is a vector density the divergence of which removes the second derivatives of $g_{\m\n}$ from $\L$:
\begin{equation}
	\hat{k}^{\m} = \sqrt{-g} \left( g^{\m\r} \D^{\l}_{\r\l} - g^{\k\l} \D^{\m}_{\k\l} \right).
\end{equation}
The conserved vectors and superpotentials can then be derived by applying Noether's method to $\L_G$. One considers an arbitrary displacement vector field $\x^{\m}$, expresses the Lie derivative $\mathcal{L}_{\x} \L_G$ as $\pd_{\m}(\L_G \x^{\m})$. Assuming then that Einstein's equations and contracted Bianchi identities are satisfied one finds by straightforward though not short calculations that there exists a conserved vector density $\hat{I}^{\a}$ equal to the divergence of a superpotential $\hat{I}^{\a\b}$, $\hat{I}^{\a} = \pd_{\b} \hat{I}^{\a\b}$, where
\begin{equation}
	\hat{I}^{\a\b} = \frac{1}{\k} \left( \nb^{[\a} \hat{\x}^{\b]} - \bn^{[\a} \hat{\x}^{\b]} + \x^{[\a} \left( \hat{k}^{\b]} - \bar{\hat{k}}^{\b]} \right) \right),
\end{equation}
where overbar represents replacing $g_{\m\n}$ by $\bar{g}_{\m\n}$, e.g., $\bar{\hat{k}}^{\a} = \sqrt{-\bar{g}} \bar{k}^{\a}$ (here, in the background $\bar{k}^{\a} = 0$). This is the KBL superpotential (see Refs. \citen{KBL} and \citen{PeKa} for details). For the purpose of our work dealing with conservation laws in theories involving not only metric but also scalar fields we shall follow the formalism of Ref. \citen{KBL} but generalized for arbitrary field variables by Petrov and Lompay.\cite{petrovmain} For a comprehensive review of \textit{nonlinear} perturbations and conservation laws on curved background in general relativity and other metric theories see Ref. \citen{petrovrev}.

As mentioned above, we shall analyze the covariant conserved currents in the scalar-tensor Horndeski theory. This is the most general scalar tensor theory leading to the covariant second order equations of motion for the scalar and metric field. It was used recently in series of papers.\cite{kobayashi1}$^-$\cite{kobayashi3} It propagates at most three degrees of freedom. Intriguingly, it is conceivable that more general scalar tensor theories exist, whose equations of motion are higher order but at the same time are characterized by constraints that remove additional, undesired, degrees of freedom. A proposal in this direction is the theory of ``beyond Horndeski'' or even ``beyond beyond Horndeski'' extensions (see, e.g., Refs. \citen{langlois1} and \citen{langlois2}, see also Ref. \citen{babichev} where black holes in the Horndeski theories and beyond are discussed). We, however, consider the original general Horndeski system only since even within this framework some very interesting models can be constructed. For example, there are special subclasses of the Horndeski theory, sometimes called Fab Four or Fab Five. These models exhibit the so-called screening property and the self-acceleration. See Ref. \citen{starobinsky} for the most recent discussion of a number of intriguing aspects of these models. We also refer the reader to some other most recent literature as, for example, Refs. \citen{CKT} and \citen{BaEsp}, where the generalized Horndeski theories are analyzed in detail from both the cosmological and local perspectives. Interestingly, the most recent work on ``Ultralight scalars as cosmological dark matter'' gives the action for the scalar field which is in the form of the Horndeski action [see equation (2) in Ref. \citen{witten} in which $a$ is replaced by $\varphi$]. By learning how to proceed to find conserved currents within the original general Horndeski framework, one should be able to tackle similarly any of these generalized theories. It is also important to notice that as a special cases of the Horndeski system one can extract any of the so-called $f(R)$ theories or the prototypes of the Brans-Dicke theory.\cite{FuMa}

Before describing how the paper is organized, let us briefly return to the issue of conserved charges and Noether currents in general relativity and its extensions. These have been constructed by using various approaches, and even in four dimensions, the expressions for the energy of Kerr-AdS black holes, for example, in general, can disagree (see Ref. \citen{BaCo}). After finishing the present work, we were informed that recently the off-shell Noether current and conserved charge in Horndeski theory were formulated in Ref. \citen{Pe} by a generalization of the Abbott-Deser-Tekin formalism.\cite{KimYi} This formalism is based on the linearized perturbations of the field equations in a fixed background satisfying the vacuum equations of motion. By contrast, as noticed above, our formalism is inspired by the KBL approach in which the perturbations of a general background spacetime $\bar{M}$ with metric $\bar{g}_{\m\n}$ may be nonlinear (large), and the Lagrangian for a perturbed system contains an appropriate divergence term (in addition to the difference of the Einstein Lagrangians for $g_{\m\n}$ and $\bar{g}_{\m\n}$), as given in Eq. (\ref{KBLLg}). It is worth noticing that under suitable conditions for Kerr-AdS black holes, the linearized KBL expression for the energy at infinity coincides with the one derived by Abbott and Deser [see Eqs. (2.9), (2.12) in Ref. \citen{BaCo}], but the coincidence is not evident. The mass of a static black hole with charge and scalar field, described by a specific Einstein-Horndeski-Maxwell theory, was expressed by the black-hole thermodynamics as developed originally by Wald (see Refs. \citen{FPoI} and \citen{FPoII}). The Noether symmetry for some special Horndeski Lagrangians was studied in the context of cosmological models recently in Refs. \citen{MM} and \citen{DG}. In contrast to the works quoted above, our contribution is general in treating non-linear perturbations with respect to general backgrounds. We thank an anonymous referee for informing us about Refs. \citen{Pe}, \citen{MM}, and \citen{DG}.

The paper is organised as follows. In Sec. II, we recall the Lagrangians of the Horndeski theory and introduce the notation. In Sec. III, the general methodology of Ref. \citen{petrovmain} is used to prepare necessary mathematical quantities for the case of theories involving both metric and scalar fields. By employing the background (auxiliary) metric $\bar{g}_{\m\n}$ the specific expressions for the conserved currents in a general Horndeski-type theory are written down at the end of the section.

The following Sec. IV has a technical character: it is shown here how to calculate the relevant expressions for current quantities. The exact forms of the conserved currents following from each type of the four Horndeski Lagrangians are derived in Sec. V. There the relevant expressions are also obtained, demonstrating how the resulting terms corresponding to the full Horndeski Lagrangian contain simpler expressions associated just with the Einstein-Hilbert action. All relevant current coefficients for the Einstein-Hilbert Lagrangian are given in the Appendix. As is the case of the KBL currents, all generalized currents for the Horndeski theory can be obtained from superpotentials (Sec. VI). Regarding the applications of the KBL formalism in, for example, a cosmological context, we study superpotentials associated with perturbations of metric and scalar fields in Sec. VII. Here, the results for linear perturbations are presented in detail. The final expressions are very lengthy; however, they considerably simplify if the background scalar field is constant. In the Sec. VIII, we specialize the expressions obtained in a general case for the Brans-Dicke theory as one of the simplest cases of the Horndeski theory involving both scalar and metric fields.

After developing the necessary mathematical tools for construction of conserved quantities in general Horndeski-type theory, the formalism can be applied for any type of combination of Lagrangians involving metric and scalar fields leading to the field equations at most of the second order. It can be employed in any specific physically motivated problem like black holes or evolution of cosmological perturbations in a Horndeski-type theory.

\section{The Horndeski scalar-tensor theory}

The Horndeski theory, the most general scalar-tensor theory of gravitation with the second-order field equations, was originally developed in Ref. \citen{horndeski} and then rederived in Ref. \citen{deffayet}. The theory with field variables being scalar field $\varphi$ and metric tensor $g_{\mu \nu}$ is given by the Lagrangian
\begin{align}
	\L &= \sqrt{-g} \sum_{i=2}^{5} \Lh_i, \label{lag}
\end{align}
where $g$ denotes the metric determinant. The individual Lagrangians in the sum are given by the following expressions
\begin{align}
	\Lh_2 &= K(\f, X), \label{lag2} \\
	\Lh_3 &= -G_3(\f, X) \Box \f, \label{lag3} \\
	\Lh_4 &= G_4(\f, X) \, R + G_{4,X} \left[ (\Box \f)^2 - \mbox{Tr} \, \Pi^2 \right], \label{lag4} \\
    \Lh_5 &= G_5(\f, X) \, G^{\m\n} \nabla_{\m\n} \f - \frac{1}{6} G_{5,X} \left[ (\Box \f)^3 - 3 \Box \f \, \mbox{Tr} \, \Pi^2 + 2 \, \mbox{Tr} \, \Pi^3 \right], \label{lag5}
\end{align}
$K$, $G_3$, $G_4$ and $G_5$ are arbitrary functions of $\varphi$ and $X$, where $X$ denotes the quadratic ``kinetic'' term, $X = - \half \partial_{\m} \f \partial^{\m} \f$, $G_{i,X} = \pd G_i / \pd X$; $R$ is the Ricci scalar and $G_{\m\n}$ denotes the Einstein tensor. The $\Box \varphi$ represents the d'Alembert operator, $\Box \varphi = g^{\m\n} \nabla_{\m\n} \varphi$. Finally, $\Pi$ is a matrix of the second covariant derivatives of $\f$, i.e. $\Pi{_{\m}}^{\n} = \nabla{_{\m}}^{\n} \f = g^{\n\r} \nabla_{\m\r} \f$; covariant derivative $\nabla_{\m}$ is associated with metric field $g_{\m\n}$ via Levi-Civita connection, in the index notation, it will be denoted by semicolon, e.g., $\f_{;\m\n}$.

\section{Conserved currents -- general expressions}

\subsection{Formulas for arbitrary Lagrangian theory}

Methods of obtaining conserved currents from expressions involving general Lagrangians are described abundantly in literature. Our aim is to derive the ``covariantized Noether identities'' by introducing an auxiliary metric. This metric, considered to be associated with a given background spacetime, provides the tool to formulate conservation laws for perturbations with respect to the background. This procedure was formulated in Ref. \citen{KBL}, for example, formalized in Refs. \citen{PeKa}, \citen{JuSi1}, and \citen{JuSi2}. The procedure was generalized to describe currents for perturbations in the Einstein-Gauss-Bonnet gravity and in the metric torsion theories of gravity.\cite{petrovEGB}$^{-}$\cite{petrovtorsion2} In this paper, we shall especially employ the work of Petrov and Lompay \cite{petrovmain} in which covariantized conservation laws are formulated for perturbations in terms of general field variables $Q_B$. In general, field variables $Q_B$ are tensor components and $B$ is an arbitrary multiindex. The field theory is described by the Lagrangian density assumed to contain fields up to their second derivatives, $\hat{L} = \hat{L}( Q_B, Q_{B,\a}, Q_{B,\a\b})$, hat denotes the scalar density. In our case, $Q_B$ represents both the metric $g_{\m\n}$ and the scalar field $\varphi$. $\hat{L}$ will be the Horndeski Lagrangian (\ref{lag}).

The first step in deriving a covariant conserved current is to introduce the auxiliary background metric $\bar{g}_{\m\n}$ into the Lagrangian. This is done by converting partial derivatives into covariant ones using $\bar{g}_{\m\n}$-Levi-Civita connection; in our case of Horndeski theory, simply by converting the $g_{\m\n}$-covariant derivatives into the $\bar{g}_{\m\n}$-covariant derivatives. These derivatives will be denoted by $\bar{\nabla}_{\a}$ or, when it is convenient, by a vertical line, e.g., $\varphi_{|\m\n}$. The new Lagrange function $\L(Q_B, \bar{\nabla}_{\a} Q_B, \bar{\nabla}_{\a\b} Q_B, \bar{g}_{\m\n}, \bar{R}{^{\l}}_{\t\r\s})$ is then obtained; $\bar{R}{^{\l}}_{\t\r\s}$ denotes the Riemann tensor formed from metric $\bar{g}_{\m\n}$.

A conserved current $\hat{i}^{\a}(\x^{\m})$ associated with an arbitrary vector field $\x^{\m}$ is given by the relation (for details, see  Ref. \citen{petrovmain}, Eqs. (37), (41), (44), and (47)) 
\begin{equation}
  \pd_{\a} \hat{i}^{\a} = \pd_{\a} \left( \hat{u}^{\a}_{\s} \xi^{\s} + \hat{m}^{\a\t}_{\s} \bar{\nabla}_{\t}\x^{\s} + \hat{n}^{\a\t\b}_{\s} \bar{\nabla}_{\t\b} \x^{\s} \right) = 0, \label{icur}
  \end{equation}
which represents the main identity of the formalism. The coefficients $\hat{u}^{\a}_{\s}$, $\hat{m}_{\s}^{\a\t}$ and $\hat{n}_{\s}^{\a\t\b}$ in (\ref{icur}) are to be calculated from the Lagrangian according to the following formulas:
\begin{align}
	\label{curn} \hat{n}^{\a\t\b}_{\s} &= \half \left[ \frac{\pd \L}{\pd Q_{B|\b\a}} \left. Q_B \right|^{\t}_{\s} + \frac{ \pd \L }{ \pd Q_{B|\t\a} } \left. Q_B \right|^{\b}_{\s} \right], \\
	\label{curm} \hat{m}^{\a\t}_{\s} &= \left[ \frac{\pd \L}{\pd Q_{B|\a}} - \bn_{\b} \left( \frac{\pd \L}{\pd Q_{B|\a\b} } \right) \right] \left. Q_B \right|^{\t}_{\s} - \frac{\pd \L}{\pd Q_{B|\t\a} } \bn_{\s} Q_B + \frac{\pd \L}{\pd Q_{B|\b\a}} \bn_{\b} \left( \left. Q_B \right|^{\t}_{\s} \right), \\
	\label{curu} \hat{u}^{\a}_{\s} &= \L \d^{\a}_{\s} + \frac{\d \L}{\d Q_B} \left. Q_B \right|^{\a}_{\s} - \left[ \frac{\pd \L}{\pd Q_{B|\a}} - \bn_{\b} \left( \frac{\pd \L}{\pd Q_{B|\a\b}} \right) \right] \bn_{\s} Q_B - \frac{\pd \L}{\pd Q_{B|\b\a}} \bn_{\b\s} Q_B \nonumber \\
	& \quad + \half \frac{\pd \L}{\pd Q_{B|\t\a}} \left. Q_B \right|^{\b}_{\l} \bar{R}{^{\l}}_{\s\t\b}.
\end{align}
The quantity $\left. Q_B \right|^{\t}_{\s}$ is defined via the Lie derivative of $Q_B$ as follows:
\begin{equation}
	\Lh_{\x} Q_B = \partial_{\r} Q_B \x^{\r} - \left. Q_B \right|^{\t}_{\s} \pd_{\t} \x^{\s}. \label{barnot}
\end{equation}

In order to find conserved currents in the Horndeski theory we need to introduce the background auxiliary metric $\bar{g}_{\m\n}$ into the Lagrangian. Such Lagrangian then enters the derivatives in (\ref{curn})--(\ref{curu}). In the following, we successively introduce the background metric into scalar field and (physical) metric field.

\subsection{The scalar field}

In Lagrangians (\ref{lag2})--(\ref{lag5}), we have the first derivatives of scalar field $\f$ in quadratic term $X = - \half \partial_{\m} \f \partial^{\m} \f$ and the second derivatives in terms $\nabla_{\m\n} \f$, $\Box \f$, $\mbox{Tr} \, \Pi^2$ and $\mbox{Tr} \, \Pi^3$. 

Since $ \pd_{\a} \f = \nabla_{\a} \f = \bar{\nabla}_{\a} \f $, we get $ X = - \half g^{\a\b} \bar{\nabla}_{\a} \f \, \bar{\nabla}_{\b} \f$. The second derivatives are replaced by 
\begin{equation}
\nabla_{\m\n} \f = \bar{\nabla}_{\m\n} \f - \D^{\r}_{\m\n} \bar{\nabla}_{\r} \f, \label{d2fcov}
\end{equation}
where $\D_{\m\n}^{\l}$ is the difference between the Christoffel symbols $\Gamma^{\l}_{\m\n}$ and $\bar{\Gamma}^{\l}_{\m\n}$ associated with $g_{\m\n}$ and $\bar{g}_{\m\n}$, respectively. It can be written in the form
\begin{equation}
	\D^{\l}_{\m\n} = \G^{\l}_{\m\n} - \bar{\G}^{\l}_{\m\n} = \half g^{\l\r} \left( \bn_{\n} g_{\r\m} + \bn_{\m} g_{\r\n} - \bn_{\r} g_{\m\n} \right). \label{chdiff}
\end{equation}
Thus we have
\begin{align}
	\Box \f = g^{\m\n} \nabla_{\m\n} \f &= g^{\m\n} \bn_{\m\n} \f - g^{\m\n} \D^{\r}_{\m\n} \bn_{\r} \f, \label{Bfcov} \\
	\Pi{_{\m}}^{\n} &= g^{\n\l} \bn_{\m\l} \f - g^{\n\l} \D^{\r}_{\l\m} \bn_{\r} \f. \label{Picov}
\end{align}

It is worth to observe that both $g_{\m\n}$ and $\bar{g}_{\m\n}$ are tensor fields which can be used to raise and lower indices and the covariant character of expressions like (\ref{Bfcov}) and (\ref{Picov}) is preserved. Whenever covariant derivatives instead of partial derivatives are employed like in the quantities $\Delta^{\l}_{\m\n}$ (\ref{chdiff}), they are constructed using the auxiliary metric $\bar{g}_{\m\n}$ and are associated with a lower case index (like in $\bn_{\n} g_{\r\m}$). However, because of the presence of various invariants formed from scalar field $\f$ by ``original'' metric $g_{\m\n}$ like in the Lagrangian (\ref{lag2})-(\ref{lag5}), it is advantageous to use this metric to raise and lower indices. This will appear in a number of lengthy formulas in the following. So, for example, in the quantity
\begin{equation}
	\nb^{\a\b} \f = g^{\a\r} g^{\b\s} \nb_{\r\s} \f, \label{graisef}
\end{equation}
where $\nb_{\r\s} \f$ is given by (\ref{d2fcov}) in terms of $\bn$ and $\D^{\l}_{\m\n}$. In still more complicated case (which will be needed for expressions originating from Lagrangians $\Lh_4$ and $\Lh_5$), the third derivatives will be required. By direct calculations we find, for example,
\begin{equation}
	\nabla{_{\a}}^{\m\n} \f = \bar{\nabla}_{\a} \nabla^{\m\n} \f + \D^{\m}_{\r\a} \nabla^{\n\r} \f + \D^{\n}_{\r\a} \nabla^{\m\r} \f, \label{derf3a}
\end{equation}
where quantities like $\nb^{\m\n} \f$ are given by (\ref{graisef}) with $\nb_{\m\n} \f$ given by (\ref{d2fcov}). In a contracted form we get
\begin{equation}
	\nabla{_{\a\r}}^{\r} \f = \bar{\nabla}_{\a} \Box \f - \D^{\k}_{\r\a} \nabla{_{\k}}^{\r} \f + \D^{\r}_{\k\a} \nabla{_{\r}}^{\k} \f, \label{derf3b}
\end{equation}
with $\Box \f$ given by expression (\ref{Bfcov}).

However, whenever expression involving, for example, operator $\bn_{\m\n}$ occurs, quantity $\bn{_{\m}}^{\n} = \bar{g}^{\n\r} \bn_{\m\r}$.

\subsection{The metric field}

The derivatives of metric field $g_{\m\n}$ are present in Lagrangians $\Lh_4$ and $\Lh_5$ [see (\ref{lag4}) and (\ref{lag5})] in the form of the Ricci scalar $R$ and the Einstein tensor $G_{\m\n}$. The covariant version of the Riemann tensor $R{^{\l}}_{\t\r\s}$, which can be found, e.g., in Ref. \citen{KBL}, is
\begin{align}
 R{^{\l}}_{\t\r\s} &= \bn_{\r} \D^{\l}_{\t\s} - \bn_{\s} \D^{\l}_{\t\r} + \D^{\l}_{\r\eta} \D^{\eta}_{\t\s} - \D^{\l}_{\eta\s} \D^{\eta}_{\t\r} + \bar{R}{^{\l}}_{\t\r\s}, \label{covriemann} \\
  &= \half g^{\l\iota} \left( \bn_{\r\s} g_{\iota\t} + \bn_{\r\t} g_{\iota\s} - \bn_{\r\iota} g_{\t\s} - \bn_{\s\r} g_{\iota\t} - \bn_{\s\t} g_{\iota\r} + \bn_{\s\iota} g_{\t\r} \right) \nonumber \\
  & \quad + Q{^{\l}}_{\t\r\s}(g_{\m\n}, \bar{\nabla}_{\a} g_{\m\n}, \bar{g}_{\m\n}, \bar{R}{^{\l}}_{\t\r\s}); \nonumber
\end{align}
here on the second line we highlighted the second derivatives of the metric. We notice that the covariantization of the Riemann tensor is ambiguous: the antisymmetric part of the second covariant derivatives can be converted into the terms involving the Riemann tensor of the auxiliary metric $\bar{R}{^{\l}}_{\t\r\s}$ multiplied by the metric $g_{\m\n}$. In this paper, we will use the covariantization in which this whole antisymmetric part is inserted into the second part $Q{^{\l}}_{\t\r\s}$, leading thus to 
\begin{align}
  R{^{\l}}_{\t\r\s} &= \half g^{\l\iota} \left( \bn_{(\r\t)} g_{\iota\s} - \bn_{(\r\iota)} g_{\t\s} - \bn_{(\s\r)} g_{\iota\t} + \bn_{(\s\iota)} g_{\t\r} \right) \nonumber \\
  & \quad + \tilde{Q}{^{\l}}_{\t\r\s}(g_{\m\n}, \bar{\nabla}_{\a} g_{\m\n}, \bar{g}_{\m\n}, \bar{R}{^{\l}}_{\t\r\s}).
\end{align}
Let us remark that this corresponds to calculating conserved current in the form $\half \hat{i}^{\a} + \half \hat{i}^{*\a}$ as denoted in Ref. \citen{petrovmain} in equation (80) and described in Section 3.2 therein as ``another variant of covariantization.''

\subsection{Conserved currents formulas in the Horndeski theory}

Since in the Horndeski theory there are two different sets of the field variables, $Q_B \equiv (\varphi, g_{\m\n})$, we split the coefficients $\hat{u}^{\a}_{\s}$, $\hat{m}_{\s}^{\a\t}$ and $\hat{n}_{\s}^{\a\t\b}$ forming the conserved current $\hat{i}^{\a}$ (\ref{icur}) into gravitational and scalar parts:
\begin{align}
	\hat{u}^{\a}_{\s} &= \L \d^{\a}_{\s} + \hat{u}^{\a}_{\s(g)} + \hat{u}^{\a}_{\s(\f)},	&	\hat{m}^{\a\t}_{\s} &= \hat{m}^{\a\t}_{\s(g)} + \hat{m}^{\a\t}_{\s(\f)},	&	\hat{n}_{\s}^{\a\t\b} &= \hat{n}_{\s(g)}^{\a\t\b} + \hat{n}_{\s(\f)}^{\a\t\b}; \label{splitcoef}
\end{align}
in the coefficient $\hat{u}^{\a}_{\s}$  we separated the $\L \d^{\a}_{\s}$ term; it will be considered later.

For the scalar part, we have $\left. \f \right|^{\t}_{\s} = 0$, hence, the formulas (\ref{curn}), (\ref{curm}) and (\ref{curu}) considerably simplify:
\begin{align}
	\hat{n}^{\a\t\b}_{\s(\f)} &= 0, \label{fcurn} \\
	\hat{m}^{\a\t}_{\s(\f)} &= - \frac{\pd \L}{\pd \f_{|\t\a}} \bn_{\s} \f, \label{fcurm} \\
	\hat{u}^{\a}_{\s(\f)} &= - \left[ \frac{\pd \L}{\pd \f_{|\a}} - \bn_{\b} \left( \frac{\pd \L}{\pd \f_{|\a\b}} \right) \right] \bn_{\s} \f - \frac{\pd \L}{\pd \f_{|\b\a}} \bn_{\b\s} \f. \label{fcuru}		
\end{align}

For the metric field $g_{\m\n}$, we have from the definition (\ref{barnot})
\begin{equation}
 \left. g_{\m\n} \right|^{\t}_{\s} = - \d^{\t}_{\m} g_{\s\n} - \d^{\t}_{\n} g_{\m\s}. \label{gbarnot}
\end{equation}
Since in (\ref{curn})--(\ref{curu}) there is always a sum over $(\m, \n)$ of $\left. g_{\m\n} \right|^{\t}_{\s}$ and a term of the type $\frac{\pd \L}{\pd g_{\m\n|\cdots}}$, we can simplify (\ref{gbarnot}) into $\left. g_{\m\n} \right|^{\t}_{\s} S^{\m\n} = - 2 g_{\r\s} S^{\t\r}$, with $S^{\m\n}$ being an arbitrary symmetric tensor. This leads to the following formulas for the conserved current coefficients:
\begin{align}
	\hat{n}^{\a\t\b}_{\s(g)} &= -g_{\r\s} \left( \frac{\pd \L}{\pd g_{\t\r|\b\a}} + \frac{\pd \L}{\pd g_{\b\r|\t\a}} \right), \label{gcurn} \\
	\hat{m}^{\a\t}_{\s(g)} &= -2g_{\r\s} \left[ \frac{\pd \L}{\pd g_{\r\t|\a}} - \bn_{\b} \left( \frac{\pd \L}{\pd g_{\r\t|\a\b}} \right) \right] - \frac{\pd \L}{\pd g_{\m\n|\t\a}} \bn_{\s} g_{\m\n} - 2 \frac{\pd \L}{\pd g_{\r\t|\b\a}} \bn_{\b} g_{\r\s}, \label{gcurm} \\
	\hat{u}^{\a}_{\s(g)} &= - 2 \frac{\d \L}{\d g_{\a\r}} g_{\s\r} - \left[ \frac{\pd \L}{\pd g_{\m\n|\a}} - \bn_{\b} \left( \frac{\pd \L}{\pd g_{\m\n|\a\b}} \right) \right] \bn_{\s} g_{\m\n} - \frac{\pd \L}{\pd g_{\m\n|\b\a}} \bn_{\b\s} g_{\m\n} \nonumber \\
	& \quad  - \frac{\pd \L}{\pd g_{\b\r|\t\a}} g_{\r\l} \bar{R}{^{\l}}_{\s\t\b}. \label{gcuru}
\end{align}

As our Lagrangian consists of four parts, we decompose the current coefficients into four parts as well:
\begin{align}
	\hat{n}^{\a\t\b}_{\s(g)} &= \sum_{i=2}^5 \hat{n}^{\a\t\b}_{\s(g)(i)}, & \hat{n}^{\a\t\b}_{\s(\f)} &= \sum_{i=2}^5 \hat{n}^{\a\t\b}_{\s(\f)(i)}, \\
	\hat{m}^{\a\t}_{\s(g)} &= \sum_{i=2}^5 \hat{m}^{\a\t}_{\s(g)(i)}, & \hat{m}^{\a\t}_{\s(\f)} &= \sum_{i=2}^5 \hat{m}^{\a\t}_{\s(\f)(i)}, \\
	\hat{u}^{\a}_{\s(g)} &= \sum_{i=2}^5 \hat{u}^{\a}_{\s(g)(i)}, & \hat{u}^{\a}_{\s(\f)} &= \sum_{i=2}^5 \hat{u}^{\a}_{\s(\f)(i)}.
\end{align}
Hereafter, this notation will be used.

\section{The derivatives of the Horndeski Lagrangian}

We give some intermediate results necessary to calculate the conserved current coefficients $\hat{u}^{\a}_{\s}$, $\hat{m}_{\s}^{\a\t}$, and $\hat{n}_{\s}^{\a\t\b}$. Simultaneously, it is rather illustrative to see how the covariantization and differentiation with respect to auxiliary fields work. As we have two different sets of fields, it is natural to divide the calculations into two parts.

\subsection{Derivatives with respect to the scalar field \label{secderf}}

The Horndeski Lagrangian consists of several constituents containing derivatives of field $\f$: the kinetic term $X$, the d'Alambertian $\Box \f$ and the trace of powers of matrix $\Pi$, $\mbox{Tr}(\Pi^k)$, $k = 2, \, 3$. For the calculation of the derivatives we need to write these terms manifestly covariant with respect to the auxiliary metric $\bar{g}_{\m\n}$; but after the calculation is done, we will eliminate the auxiliary covariant differentiation $\bar{\nabla}_{\a}$ in favor of covariant derivative $\nabla_{a}$ and Christoffel symbols difference $\Delta^{\l}_{\m\n}$ whenever possible.

The quadratic term $X$ is simple: $ X = -\half g^{\m\n} \nabla_{\m} \f \nabla_{\n} \f = -\half g^{\m\n} \bar{\nabla}_{\m} \f \bar{\nabla}_{\n} \f$. Hence,
\begin{align}
	\frac{\pd X}{\pd \f_{|\a}} &= - g^{\a\r} \nabla_{\r} \f = - \nabla^{\a} \f, & \frac{\pd X}{\pd \f_{|\a\b}} &= 0.
\end{align}
Notice that for raising and lowering indices only the metric field $g_{\m\n}$ is always used.

The derivatives of $\nabla_{\m\n} \f$ are easily found employing (\ref{d2fcov}):
\begin{align}
	\frac{\pd \f_{;\m\n}}{\pd \f_{|\a}} &= - \D^{\a}_{\m\n}, & \frac{\pd \f_{;\m\n}}{\pd \f_{|\a\b}} &= \d^{(\a}_{(\m} \d^{\b)}_{\n)}, \label{derd2f}
\end{align}
which are in turn used to calculate all necessary terms present in the Horndeski Lagrangian, i.e. the d'Alambertian and its powers:
\begin{align}
	\frac{\pd \, \Box \f}{\pd \f_{|\a}} &= - g^{\m\n} \D^{\a}_{\m\n}, & \frac{\pd \, \Box \f}{\pd \f_{|\a\b}} &= g^{\a\b}, \label{derboxf} \\
	\frac{\pd (\Box \f)^k}{\pd \f_{|\a}} &= - k (\Box \f)^{k-1} g^{\m\n} \D^{\a}_{\m\n} , & \frac{\pd (\Box \f)^k}{\pd \f_{|\a\b}} &= k (\Box \f)^{k-1} g^{\a\b}. \label{derboxfk}
\end{align}
The matrix of the second derivatives $\Pi{^{\m}}_{\n}$ of the scalar field $\f$ and the trace of its various powers are given by
\begin{align}
	\frac{\pd \Pi{_{\m}}^{\n}}{\pd \f_{|\a}} &= - g^{\n\l} \D^{\a}_{\l\m}, & \frac{\pd \Pi{_{\m}}^{\n}}{\pd \f_{|\a\b}} &= \d^{(\a}_{\m} g^{\b)\n}, \\
	\frac{\pd \, \mbox{Tr} \, \Pi^k}{\pd \f_{|\a}} &= k \, \mbox{Tr} \left( \frac{\pd \Pi}{\pd \f_{|\a}} \cdot \Pi^{k-1} \right), & \frac{\pd \, \mbox{Tr} \, \Pi^k}{\pd \f_{|\a\b}} &= k \, \mbox{Tr} \left( \frac{\pd \Pi}{\pd \f_{|\a\b}} \cdot \Pi^{k-1} \right), \\
	\frac{\pd \, \mbox{Tr} \, \Pi^2}{\pd \f_{|\a}} &= - 2 \D^{\a}_{\m\n} \nabla^{\m\n} \f, & 	\frac{\pd \, \mbox{Tr} \, \Pi^2}{\pd \f_{|\a\b}} &= 2 \nabla^{\a\b} \f, \label{dertrp2} \\
	\frac{\pd \, \mbox{Tr} \, \Pi^3}{\pd \f_{|\a}} &= - 3 \D^{\a}_{\r\s} \nb^{\l\r} \f \nb{^{\s}}_{\l} \f, & \frac{\pd \, \mbox{Tr} \, \Pi^3}{\pd \f_{|\a\b}} &= 3 \nb^{\a\r} \f \nb{^{\b}}_{\r} \f. \label{dertrp3}
\end{align}

\subsection{Derivatives with respect to the metric field $g_{\m\n}$ \label{secderg}}

The metric field $g_{\m\n}$ is explicitly present in the Ricci scalar and the Riemann tensor, and its first derivatives are also contained in the second derivatives of the scalar field $\f$, in the expression for the Christoffel symbols difference [see (\ref{d2fcov}) and (\ref{chdiff})]. Hence, using
\begin{equation}
	\frac{\pd \D^{\l}_{\t\s}}{\pd g_{\m\n|\a}} = \d^{\a}_{(\t} \d^{(\m}_{\s)} g^{\n)\l} - \half \d^{(\m}_{(\t} \d^{\n)}_{\s)} g^{\a\l},
\end{equation}
we find the following relation:
\begin{align}
	\frac{\pd \f_{;\k\l}}{\pd g_{\m\n|\a}} &= - \frac{\pd \D^{\r}_{\k\l}}{\pd g_{\m\n|\a}} \bn_{\r} \f = - \half \left( g^{\r(\m} \d^{\n)}_{\k} \d^{\a}_{\l} + g^{\r(\m} \d^{\n)}_{\l} \d^{\a}_{\k} - g^{\r\a} \d^{(\m}_{\k} \d^{\n)}_{\l} \right) \nb_{\r} \f.
\end{align}
Consequently, the resulting derivatives for the d'Alambertians look as follows:
\begin{align}
	\frac{\pd \Box \f}{\pd g_{\m\n|\a}} &= g^{\k\l} \frac{\pd \f_{;\k\l}}{\pd g_{\m\n|\a}} = \left( \half g^{\a\r} g^{\m\n} - g^{\r(\m} g^{\n)\a} \right) \nb_{\r} \f = \half g^{\m\n} \nb^{\a} \f - g^{\a(\m} \nb^{\n)} \f, \label{gderboxf} \\
		\frac{\pd (\Box \f)^k}{\pd g_{\m\n|\a}} &= k (\Box \f)^{k-1} \left( \half g^{\m\n} \nb^{\a} \f - g^{\a(\m} \nb^{\n)} \f \right), \label{gderboxfk}.
\end{align}
Finally, the traces of the second derivatives read
\begin{align}
	\frac{\pd \Pi{_{\k}}^{\l}}{\pd g_{\m\n|\a}} &= - g^{\l\iota} \frac{\pd \D^{\r}_{\iota\k}}{\pd g_{\m\n|\a}} \bn_{\r} \f = - \half \left( g^{\l(\m} g^{\n)\r} \d^{\a}_{\k} + g^{\l\a} g^{\r(\m} \d^{\n)}_{\k} - g^{\r\a} g^{\l(\m} \d^{\n)}_{\k} \right) \nb_{\r} \f, \\
	\frac{\pd \, \mbox{Tr} \, \Pi^k}{\pd g_{\m\n|\a}} &= k \, \mbox{Tr} \, \left( \frac{\pd \Pi}{\pd g_{\m\n|\a}} \cdot \Pi^{k-1} \right), \\
	\frac{\pd \, \mbox{Tr} \, \Pi^2}{\pd g_{\m\n|\a}} &= \nb^{\a} \f \nb^{\m\n} \f - 2 \nb^{(\m} \f \nb^{\n)\a} \f, \label{gdertrp2} \\
	\frac{\pd \, \mbox{Tr} \, \Pi^3}{\pd g_{\m\n|\a}} &= \frac{3}{2} \left( \nb^{\a} \f \nb^{\r(\m} \f \nb{^{\n)}}_{\r} \f - 2 \nb^{(\m} \f \nb^{\n)\r} \f \nb{_{\r}}^{\a} \f \right). \label{gdertrp3}
\end{align}

For the derivatives of the Riemann tensor which are necessary in the calculation of the conserved current coefficients for Lagrangians $\Lh_4$ and $\Lh_5$ see Secs. \ref{coefl4} and \ref{coefl5} and the Appendix.

\section{Conserved current coefficients}

\subsection{Conserved current coefficients for $\Lh_2$ \label{coefl2}}

Starting from (\ref{lag2}) and employing the preceding results, we easily obtain the conserved current coefficients $\hat{u}^{\a}_{\s(2)}$, $\hat{m}_{\s(2)}^{\a\t}$ and $\hat{n}_{\s(2)}^{\a\t\b}$. First, derivatives with respect to the scalar field derivatives are
\begin{align}
	\frac{\pd \Lh_2}{\pd \f_{|\a}} &= - \pd_X K \, \nb^{\a} \f, & \frac{\pd \Lh_2}{\pd \f_{|\a\b}} &= 0,
\end{align}
which leads to $\hat{m}^{\a\t}_{\s(\f)(2)} = 0$ and $\hat{u}^{\a}_{\s(\f)(2)} = \pd_{X} \hat{K} \, \nabla^{\a} \f \nabla_{\s} \f$. Derivatives with respect to the metric field $g_{\m\n}$ all vanish for $\Lh_2$ and, consequently, the coefficients $\hat{m}_{\s(g)(2)}^{\a\t}$ and $\hat{n}_{\s(g)(2)}^{\a\t\b}$ vanish as well; only in $\hat{u}^{\a}_{\s(g)(2)}$ the contribution from the field equations remains:
\begin{equation}
	\hat{u}^{\a}_{\s(g)(2)} = - 2 \frac{\d \L_2}{\d g_{\a\r}} g_{\r\s}.
\end{equation}

The total conserved current coefficients for the Lagrangian $\Lh_2$ are thus given by
\begin{align}
	\hat{u}^{\a}_{\s(2)} &= \L_2 \d^{\a}_{\s} + \pd_{X} \hat{K} \, \nabla^{\a} \f \nabla_{\s} \f - 2 \frac{\d \L_2}{\d g_{\a\r}} g_{\r\s}, & \hat{m}_{\s(2)}^{\a\t} &= 0, & \hat{n}_{\s(2)}^{\a\t\b} = 0.
\end{align}

\subsection{Conserved current coefficients for $\Lh_3$ \label{coefl3}}

The Horndeski Lagrangian $\Lh_3$ contains the second covariant derivatives of the scalar field. These have  an obvious impact on the derivative of $\Lh_3$ w.r.t. the second partial derivative of the field $\f$, but they also imply the non-vanishing derivatives w.r.t. the first derivatives of the field and metric $g_{\m\n}$ [see (\ref{d2fcov}) and (\ref{derboxf})].

For the scalar part we get
\begin{align}
	\frac{\pd \Lh_3}{\pd \f_{|\a}} &= \pd_{X} G_3 \, \nabla^{\a} \f \, \Box \f + G_3 g^{\m\n} \D^{\a}_{\m\n}, \label{dL3dnf} \\
	\frac{\pd \Lh_3}{\pd \f_{|\a\b}} &= -G_3 \, g^{\a\b}.
\end{align}
It is simple to insert these expressions into (\ref{fcurm}) and (\ref{fcuru}) to obtain coefficients $\hat{m}^{\a\t}_{\s(\f)(3)}$ and $\hat{u}^{\a}_{\s(\f)(3)}$. However, there is one term involved in the calculation of $\hat{u}^{\a}_{\s(\f)(3)}$ which needs to be examined more closely:
\begin{equation}
	\bar{\nabla}_{\b} \left( \frac{\pd \L_3}{\pd \f_{|\a\b}} \right) = - \bar{\nabla}_{\b} \left( \sqrt{-g} \, G_3(\f, X) g^{\a\b} \right).
\end{equation}
As noticed already, the result should be written in terms of $g$-covariant derivatives and the Christoffel symbols difference $\D^{\l}_{\m\n}$. For the derivative of the metric and its determinant, we have
\begin{equation}
	\bar{\nabla}_{\a} g^{\m\n} = - 2 g^{\l(\m} \D^{\n)}_{\l\a}, \qquad \bar{\nabla}_{\a} g_{\m\n} = 2 g_{\l(\m} \D^{\l}_{\n)\a};
\end{equation}
and $ \bar{\nabla}_{\a} g = 2 g \D^{\l}_{\l\a} $, since the metric determinant is the scalar density of weight two, implying
\begin{equation}
	\bar{\nabla}_{\a} \sqrt{-g} = \sqrt{-g} \, \D^{\l}_{\l\a}. \label{dsqrtdetg}
\end{equation}
The derivative of $G_3$, a function of two scalars, $\bar{\nabla}_{\a} \f = \nabla_{\a} \f$, $\bar{\nabla}_{\a} X = \nabla_{\a} X = - \nabla_{\a\r} \f \nabla^{\r} \f$, is
\begin{equation}
	\bar{\nabla}_{\a} G_3 = \pd_{\f} G_3 \nabla_{\a} \f - \pd_{X} G_3 \nabla_{\a\r} \f \nabla^{\r} \f. \label{bnG3}
\end{equation}

Regarding (\ref{dL3dnf})--(\ref{bnG3}), we find the final forms of $\hat{m}_{\s(\f)(3)}^{\a\t}$ and $\hat{u}^{\a}_{\s(\f)(3)}$:
\begin{align}
	\hat{m}^{\a\t}_{\s(\f)(3)} &= \hat{G}_3 g^{\t\a} \nabla_{\s} \f, \\
	\hat{u}^{\a}_{\s(\f)(3)} &= \hat{G}_3 ( \nabla{^{\a}}_{\s} \f + g^{\a\k} \D^{\r}_{\s\k} \nabla_{\r} \f ) - \pd_{\f} \hat{G}_3 \nabla^{\a} \f  \nabla_{\s} \f \nonumber \\
		& \quad + \pd_X \hat{G}_3 \left( \nabla^{\a\r} \f \nabla_{\r} \f \nabla_{\s} \f - \Box \f \nabla^{\a} \f \nabla_{\s} \f \right).
\end{align}

For the metric part, the results considerably simplify since $\Lh_3$ does not contain the second derivatives of $g_{\m\n}$. The non-vanishing derivatives of $\Lh_3$ are obtained using (\ref{gderboxf}). The result is
\begin{align}
	\frac{\pd \Lh_3}{\pd g_{\m\n|\a}} &= G_3 \left( g^{\a(\m} \nabla^{\n)} \f - \half g^{\m\n} \nabla^{\a} \f \right), \\
	\frac{\pd \Lh_3}{\pd g_{\m\n|\a\b}} &= 0.
\end{align}
The above expressions plugged into formulas (\ref{gcurn})--(\ref{gcuru}) lead to the following conserved current coefficients for $\Lh_3$:
\begin{align}
	\hat{n}^{\a\t\b}_{\s(g)(3)} &= 0, \label{l3gn} \\
	\hat{m}^{\a\t}_{\s(g)(3)} &= -2g_{\r\s} \frac{\pd \L_3}{\pd g_{\r\t|\a}} = \hat{G}_3 \left( 2 \d_{\s}^{[\t} \nabla^{\a]} \f - g^{\a\t} \nabla_{\s} \f \right), \label{l3gm} \\
	\hat{u}^{\a}_{\s(g)(3)} &= - 2 \frac{\d \L_3}{\d g_{\a\r}} g_{\s\r} - \frac{\pd \L_3}{\pd g_{\m\n|\a}} \bn_{\s} g_{\m\n} \nonumber \\
	 	&= - 2 \frac{\d \L_3}{\d g_{\a\r}} g_{\s\r} + \hat{G}_3 \left( \nabla^{\a} \f \D^{\r}_{\r\s} - \nabla^{\r} \f \D^{\a}_{\r\s} - \nabla_{\r} \f \D^{\r}_{\k\s} g^{\k\a} \right). \label{l3gu}
\end{align}

\subsection{Conserved current coefficients for $\Lh_4$ \label{coefl4}}

The Lagrangian $\Lh_4$ contains the scalar curvature $R$ which, multiplied by the square root of the determinant $g$, constitutes the Einstein-Hilbert Lagrangian. The calculations for conserved current coefficients will be performed in such a way that the coefficients corresponding to the Einstein-Hilbert Lagrangian will be preserved in the final result.

The derivatives of $\Lh_4$ with respect to the scalar field derivatives are easily obtained using expressions (\ref{derboxfk}) and (\ref{dertrp2}):
\begin{align}
	\frac{\pd \Lh_4}{\pd \f_{|\a}} &= - \pd_X G_4 \, \nb^{\a} \f \, R - \pd_{XX} G_4 \, \nabla^{\a} \f \left[ (\Box \f)^2 - \mbox{Tr} \, \Pi^2 \right] - 2 \pd_{X} G_4 \, \D^{\a}_{\m\n} \left[ g^{\m\n} \Box \f - \nb^{\m\n} \f \right], \label{l4df1} \\
	\frac{\pd \Lh_4}{\pd \f_{|\a\b}} &= \pd_{X} G_4 \left[ 2 \Box \f \, g^{\a\b} - 2 \nb^{\a\b} \f \right]. \label{l4df2}
\end{align}
To calculate the conserved current, we plug the above results into the formulas (\ref{fcurm}) and (\ref{fcuru}). The calculations are straightforward except, as in the case of $\Lh_3$, for the term 
\begin{equation}
	\bar{\nabla}_{\b} \left( \frac{\pd \L_4}{\pd \f_{|\a\b}} \right) = \D^{\l}_{\l\b} \frac{\pd \L_4}{\pd \f_{|\a\b}} + \sqrt{-g} \, \bar{\nabla}_{\b} \left( \frac{\pd \Lh_4}{\pd \f_{|\a\b}} \right); \label{l4hatterm}
\end{equation}
we used the relation $\bar{\nabla}_{\a}( \sqrt{-g} \, T ) = \D^{\l}_{\l\a} \hat{T} + \sqrt{-g} \, \bar{\nabla}_{\a} T$, where $T$ denotes an arbitrary tensor, c.f. (\ref{dsqrtdetg}). Furthermore, in the second term of (\ref{l4hatterm}), we need to convert the third ``mixed'' covariant derivatives $\bar{\nabla}_{\a} \Box \f$ and $\bar{\nabla}_{\a} \nabla^{\m\n} \f$ into just the $g$-covariant derivatives and $\D^{\a}_{\m\n}$-s using formulas (\ref{derf3a}) and (\ref{derf3b}).
The resulting conserved current coefficients for scalar part then read
\begin{align}
	\hat{m}^{\a\t}_{\s(\f)(4)} &= - \pd_X \hat{G}_4 \left[ 2 \Box \f \, g^{\t\a} - 2 \nb^{\a\t} \f \right] \nb_{\s} \f, \\
	\hat{u}^{\a}_{\s(\f)(4)} =& \, \pd_{X} \hat{G}_4 \left( R \nabla_{\s} \f \nabla^{\a} \f + 2 \left( \nabla_{\s\r} \f \nabla^{\r\a} \f - \Box \f \nabla{_{\s}}^{\a} \f \right) \right. \nonumber \\
	& \, \qquad \left. + 2 \D^{\r}_{\k\s} \nabla_{\r} \f \left( \nabla^{\k\a} \f - g^{\k\a} \Box \f \right) + 2 \nabla_{\s} \f \left( \nabla^{\a}(\Box \f) - \nabla{_{\r}}^{\a\r} \f \right)  \right) \nonumber \\
		& \, + \pd_{XX} \hat{G}_4 \nabla_{\s} \f \left( \nabla^{\a} \f \left( (\Box \f)^2 - \mbox{Tr} \, \Pi^2 \right) + 2 \nabla_{\r} \f \left( \nabla^{\r\k} \f \nabla{_{\k}}^{\a} \f - \Box \f \nabla^{\r\a} \f \right) \right) \nonumber \\
		& \, + 2 \pd_{X\f} \hat{G}_4 \nabla_{\s} \f \left( \Box \f \nabla^{\a} \f - \nabla^{\a\r} \f \nabla_{\r} \f \right).
\end{align}

The Lagrangian $\Lh_4$ is the first one containing the second derivatives of the metric in the scalar curvature. For now, we keep the derivatives of the scalar curvature unevaluated as it will mainly contribute to the Einstein-Hilbert part of the conserved current coefficients. Employing (\ref{gderboxfk}) and (\ref{gdertrp2}), we obtain the Lagrangian derivatives with respect to the metric field derivatives as follows:
\begin{align}
	\frac{\pd \Lh_4}{\pd g_{\m\n|\a}} &= G_4 \frac{\pd R}{\pd g_{\m\n|\a}} + \pd_X G_4 \left[ g^{\m\n} \nb^{\a} \f \, \Box \f - 2 g^{\a(\m} \nb^{\n)} \f \, \Box \f - \nb^{\a} \f \nb^{\m\n} \f + 2 \nb^{(\m} \f \nb^{\n)\a} \f \right], \label{l4dg1} \\
	\frac{\pd \Lh_4}{\pd g_{\m\n|\a\b}} &= G_4 \frac{\pd R}{\pd g_{\m\n|\a\b}}. \label{l4dg2}
\end{align}

In the conserved current, the coefficient $\hat{n}^{\a\t\b}_{\s(g)(4)}$ is simple since the result is just the Einstein-Hilbert current multiplied by the function $G_4$:
\begin{equation}
	\hat{n}^{\a\t\b}_{\s(g)(4)} = - G_4 \, g_{\r\s} \left( \frac{\pd \hat{R}}{\pd g_{\t\r|\b\a}} + \frac{\pd \hat{R}}{\pd g_{\b\r|\t\a}} \right) = G_4 \, \hat{n}^{\a\t\b}_{\s(EH)}.
\end{equation}

The next term we have to evaluate, 
\begin{align}
	\bar{\nabla}_{\b} \left( \frac{\pd \L_4}{\pd g_{\m\n|\a\b}} \right) &= \bar{\nabla}_{\b} G_4 \frac{\pd \hat{R}}{\pd g_{\m\n|\a\b}} + G_4 \bar{\nabla}_{\b} \left( \frac{\pd \hat{R}}{\pd g_{\m\n|\a\b}} \right), \label{l4hatterm2}
\end{align}
consists of two parts. The second part will contribute to the Einstein-Hilbert coefficient $\hat{m}^{\a\t}_{\s(EH)}$, for the first part we need an explicit expression for scalar curvature derivative,
\begin{align}
	\frac{\pd R}{\pd g_{\m\n|\a\b}} &= g^{\a(\m} g^{\n)\b} - g^{\a\b} g^{\m\n}. \label{rd2}
\end{align}
Using expressions (\ref{l4dg1})--(\ref{rd2}), we obtain the final conserved current coefficient $\hat{m}^{\a\t}_{\s(g)(4)}$:
\begin{align}
	\hat{m}^{\a\t}_{\s(g)(4)} =& \, G_4 \hat{m}^{\a\t}_{\s(EH)} + \pd_{\f} \hat{G}_4 \left( \nabla_{\s} \f \, g^{\a\t} + \nabla^{\t} \f \, \d^{\a}_{\s} - 2 \nabla^{\a} \f \, \d^{\t}_{\s} \right) \nonumber \\
		& \, + \pd_{X} \hat{G}_4 \Big( 2 \Box \f \left( 2 \d^{[\a}_{\s} \nabla^{\t]} \f + \nabla_{\s} \f \, g^{\a\t} \right) + 2 \left( 2 \nabla^{[\a} \f \nabla{^{\t]}}_{\s} \f - \nabla_{\s} \f \nabla^{\a\t} \f \right) \nonumber \\
		& \, \qquad + \nabla_{\r} \f \left( 2 \nabla^{\r\a} \f \, \d^{\t}_{\s} - \nabla^{\r\t} \f \, \d^{\a}_{\s} - \nabla{^{\r}}_{\s} \f \, g^{\a\t} \right) \Big)
\end{align}

By similar procedure we find
\begin{align}
	\hat{u}^{\a}_{\s(g)(4)} =& \, - 2 \frac{\d \L_4}{\d g_{\a\r}} g_{\s\r} + G_4 \left( \hat{u}^{\a}_{\s(EH)} + 2 \frac{\d \hat{R}}{\d g_{\a\r}} g_{\s\r} \right) \nonumber \\
		& \, + \pd_{\f} \hat{G}_4 \left( \D^{\a}_{\s\r} \nabla^{\r} \f + \D^{\r}_{\s\k} \nabla_{\r} \f g^{\k\a} - 2 \D^{\r}_{\r\s} \nabla^{\a} \f \right) \nonumber \\
		& \, + \pd_X \hat{G}_4 \left( 2 \Box \f \left( \D^{\a}_{\s\r} \nabla^{\r} \f + \D^{\r}_{\s\k} \nabla_{\r} \f g^{\k\a} - \D^{\r}_{\r\s} \nabla^{\a} \f \right) + 2 \D^{\r}_{\k\s} \nabla^{\a} \f \nabla{_{\r}}^{\k} \f \right. \nonumber \\
		& \quad \, - 2 \D^{\r}_{\s\k} \nabla_{\r} \f \nabla^{\k\a} \f - 2 \D^{\r}_{\s\k} \nabla^{\k} \f \nabla{_{\r}}^{\a} \f + 2 \D^{\r}_{\r\s} \nabla_{\k} \f \nabla^{\k\a} \f - \D^{\a}_{\s\r} \nabla_{\k} \f \nabla^{\r\k} \f \nonumber \\
		& \quad \, \left. - \D^{\r}_{\s\k} \nabla_{\l} \f \nabla{^{\l}}_{\r} \f g^{\k\a} \right).
\end{align}

\subsection{Conserved current coefficients for $\Lh_5$ \label{coefl5}}

The last Horndeski Lagrangian $\Lh_5$ is the most complex -- it contains the Einstein tensor $G_{\m\n}$ and the cubic terms of the second derivatives of the scalar field. The derivatives with respect to the scalar field are found rather easily using (\ref{derd2f}), (\ref{derboxfk}), (\ref{dertrp2}) and (\ref{dertrp3}):
\begin{align}
	\frac{\pd \Lh_5}{\pd \f_{|\a}} =& \, - G_5 G^{\m\n} \D^{\a}_{\m\n} + \pd_X G_5 \bigg( - G^{\m\n} \nabla^{\a} \f \nabla_{\m\n} \f + \half g^{\m\n} \D^{\a}_{\m\n} \left( (\Box \f)^2 - \mbox{Tr} \Pi^2 \right)  \nonumber \\
		& \,  - \Box \f \, \D^{\a}_{\m\n} \nabla^{\m\n} \f + \D^{\a}_{\m\n} \nabla^{\m\l} \f \nabla{_{\l}}^{\n} \f \bigg) \nonumber \\
		& \, + \pd_{XX} G_5 \nabla^{\a} \f \left( \frac{1}{6} (\Box \f)^3 - \half \Box \f \, \mbox{Tr} \Pi^2 + \frac{1}{3} \mbox{Tr} \Pi^3 \right), \label{l5df1} \\
	\frac{\pd \Lh_5}{\pd \f_{|\a\b}} =& \, G_5 G^{\a\b} + \pd_X G_5 \left( \Box \f \, \nabla^{\a\b} \f - \nabla^{\a\r} \f \nabla{_{\r}}^{\b} \f - \half g^{\a\b} \left( (\Box \f)^2 - \mbox{Tr} \Pi^2 \right) \right). \label{l5df2}
\end{align}
The conserved current coefficient $\hat{m}^{\a\t}_{\s(\f)(5)}$ is then simply obtained by putting (\ref{l5df2}) into (\ref{fcurm}):
\begin{align}
	\hat{m}^{\a\t}_{\s(\f)(5)} &= \left( \frac{1}{6} \pd_X \hat{G}_5 \left[ 3 g^{\a\t} (\Box \f)^2 - 3 g^{\a\t} \, \mbox{Tr} \, \Pi^2 - 6 \Box \f \, \nb^{\a\t} \f + 6 \nb^{\t\r} \f \nb{_{\r}}^{\a} \f \right] - \hat{G}_5 G^{\a\t} \right) \nb_{\s} \f.
\end{align}

For the term $\bar{\nabla}_{\b} ( \pd \L_5 / \pd \f_{|\a\b} )$ in $\hat{u}^{\a}_{\s(\f)(5)}$ we use (\ref{derf3a}) and (\ref{derf3b}) as in the conserved current coefficients for $\Lh_4$. Also, we use the following relation:
\begin{equation}
	\bar{\nabla}_{\b} G^{\a\b} = - \left( \D^{\a}_{\r\b} G^{\r\b} + \D^{\b}_{\r\b} G^{\a\r} \right).
\end{equation}
After quite tedious calculations we get the final form of the conserved current coefficient $\hat{u}^{\a}_{\s(\f)(5)}$ as follows:
\begin{align}
	\hat{u}^{\a}_{\s(\f)(5)} =& \, - \hat{G}_5 G^{\a\r} \left( \nabla_{\r\s} \f + \D^{\k}_{\r\s} \nabla_{\k} \f \right) + \pd_{\f} \hat{G}_5 G^{\a\r} \nabla_{\r} \f \nabla_{\s} \f \nonumber \\
		& \, + \pd_{X} \hat{G}_5 \left( \nabla_{\s} \f \left.\Big( \left( G^{\r\k} \nabla^{\a} \f - G^{\a\r} \nabla^{\k} \f \right) \nabla_{\r\k} \f  + \half \nabla^{\a} \left( \mbox{Tr} \Pi^2 - (\Box \f)^2 \right) \right. \right. \nonumber \\
		& \, \qquad \left. + \nabla^{\a\r} \f \nabla_{\r}( \Box \f ) + \Box \f \nabla{_{\t}}^{\a\t} \f - \nabla_{\r\k} \f \nabla^{\r\a\k} \f - \nabla^{\a\r} \f \nabla{_{\k\r}}^{\k} \f \right.\Big) \nonumber \\		
		& \, \qquad + \left. \left( \nabla_{\r\s} \f + \D^{\k}_{\r\s} \nabla_{\k} \f \right) \left( \half g^{\a\r} \left( (\Box \f)^2 - \mbox{Tr} \Pi^2 \right) - \Box \f \nabla^{\a\r} \f + \nabla^{\r\l} \f \nabla{_{\l}}^{\a} \f \right) \right) \nonumber \\
		& \, + \pd_{X\f} \hat{G}_5 \nabla_{\s} \f \left( \half \left( \mbox{Tr} \Pi^2 - (\Box \f)^2 \right) \nabla^{\a} \f +  \nabla_{\r} \f \left( \nabla^{\r\a} \f \, \Box \f  - \nabla^{\r\k} \f \nabla{_{\k}}^{\a} \f \right) \right) \nonumber \\
		& \, + \pd_{XX} \hat{G}_5 \nabla_{\s} \f \left( \nabla^{\a} \f \left( -\frac{1}{6} (\Box \f)^3 + \half \mbox{Tr} \Pi^2 \Box \f - \frac{1}{3} \mbox{Tr} \Pi^3 \right) \right. \nonumber \\
		& \, \qquad \left. + \nabla_{\r} \f \left( \half \left( (\Box \f)^2 - \mbox{Tr} \Pi^2 \right) \nabla^{\r\a} \f - \Box \f \nabla^{\r\k} \f \nabla{_{\k}}^{\a} \f + \nabla^{\r\k} \f \nabla_{\k\l} \f \nabla^{\l\a} \f \right) \right).
\end{align}

Now let us turn to the metric field. The derivative of $\Lh_5$ with respect to the first derivative of $g_{\m\n}$ is worked out using expressions (\ref{gderboxfk}), (\ref{gdertrp2}), and (\ref{gdertrp3}):
\begin{align}
	\frac{\pd \Lh_5}{\pd g_{\m\n|\a}} =& \, G_5 \left( \frac{\pd R^{\r\k}}{\pd g_{\m\n|\a}} \nabla_{\r\k} \f - \half \frac{\pd R}{\pd g_{\m\n|\a}} \Box \f + \half G^{\m\n} \nabla^{\a} \f - G^{\a(\m} \nabla^{\n)} \f \right) \nonumber \\
		& \, + \pd_X G_5 \left( \half \left( (\Box \f)^2 - \mbox{Tr} \Pi^2 \right) \left( g^{\a(\m} \nabla^{\n)} \f - \half g^{\m\n} \nabla^{\a} \f \right) + \nabla^{(\m} \f \nabla^{\n)\r} \f \nabla{_{\r}}^{\a} \f \right. \nonumber \\
		& \, \qquad \left. + \half \Box \f \left( \nabla^{\a} \f \nabla^{\m\n} \f - 2 \nabla^{(\m} \f \nabla^{\n)\a} \f \right) - \half \nabla^{\a} \f \nabla^{\r(\m} \f \nabla{^{\n)}}_{\r} \f  \right) \label{l5dg1},
\end{align}
where, when convenient, the Einstein tensor has been split into the Ricci tensor and the scalar curvature. The scalar curvature will then contribute to the Einstein-Hilbert part of the resulting conserved current coefficients. Using the derivative of Ricci tensor with respect to the first derivative of the metric,
\begin{align}
	\frac{\pd R_{\t\s}}{\pd g_{\m\n|\a}} =& \, \half \D^{\a}_{\r\k} \d^{(\m}_{(\t} \d^{\n)}_{\s)} g^{\r\k} - \D^{\a}_{\r(\t} \d^{(\m}_{\s)} g^{\n)\r} + \half \D^{\a}_{\t\s} g^{\m\n} - \D^{(\m}_{\r\k} \d^{\n)}_{(\t} \d^{\a}_{\s)} g^{\r\k} \nonumber \\
		& \, + \d^{\a}_{(\t} \D^{(\m}_{\s)\r} g^{\n)\r} + \D^{(\m}_{\r(\t} \d^{\n)}_{\s)} g^{\r\a} - \D^{(\m}_{\t\s} g^{\n)\a}, \label{ricd1}	
\end{align}
we express the first term of (\ref{l5dg1}) as follows:
\begin{align}
	\frac{\pd R^{\r\k}}{\pd g_{\m\n|\a}} \nabla_{\r\k} \f =& \, \half \D^{\a}_{\r\k} \nabla^{\m\n} \f \, g^{\r\k} - \D^{\a}_{\r\k} \nabla^{\r(\m} \f \, g^{\n)\k} + \half \D^{\a}_{\r\k} \nabla^{\r\k} \f \, g^{\m\n} - \D^{(\m}_{\r\k} \nabla^{\n)\a} \f \, g^{\r\k} \nonumber \\
		& \, + g^{\k(\m} \D^{\n)}_{\k\r} \nabla^{\a\r} \f + \D^{(\m}_{\r\k} \nabla^{\n)\r} \f \, g^{\a\k} - g^{\a(\m} \D^{\n)}_{\r\k} \nabla^{\r\k} \f. \label{ricfdg1}
\end{align}

The only object in $\Lh_5$ containing the second derivatives of the metric tensor is the Einstein tensor $G_{\m\n}$; we get
\begin{align}
	\frac{\pd \Lh_5}{\pd g_{\m\n|\a\b}} &= G_5 \frac{\pd R^{\r\s}}{\pd g_{\m\n|\a\b}} \nb_{\r\s} \f - \half G_5 \Box \f \frac{\pd R}{\pd g_{\m\n|\a\b}}, \label{l5dg2}
\end{align}
where, as in (\ref{l5dg1}), we split the Einstein tensor into the Ricci and scalar curvature part. The differentiated Ricci tensor with respect to the second derivatives of the metric reads
\begin{align}
	\frac{\pd R_{\t\s}}{\pd g_{\m\n|\a\b}} =& \, \half \left( \d^{\a}_{(\t} \d^{(\m}_{\s)} g^{\n)\b} + \d^{\b}_{(\t} \d^{(\m}_{\s)} g^{\n)\a} - \d^{(\a}_{(\t} \d^{\b)}_{\s)} g^{\m\n} - \d^{(\m}_{(\t} \d^{\n)}_{\s)} g^{\a\b} \right).
\end{align}
After contracting it with the second derivatives of the scalar field we get
\begin{align}
	\frac{\pd R^{\r\k}}{\pd g_{\m\n|\a\b}} \nabla_{\r\k} \f =& \, \half \left( g^{\a(\m} \nabla^{\n)\b} \f + g^{\b(\m} \nabla^{\n)\a} \f - g^{\m\n} \nabla^{\a\b} \f - g^{\a\b} \nabla^{\m\n} \f \right), \label{ricfdg2}
\end{align}
which is then inserted into the expression (\ref{l5dg2}).

Finally, we can calculate the last set of conserved current coefficients by employing the above results (\ref{l5dg1}), (\ref{l5dg2}) [together with (\ref{ricfdg1}) and (\ref{ricfdg2})]:
\begin{equation}
	\hat{n}^{\a\t\b}_{\s(g)(5)} = - \half \hat{G}_5 \left( \Box \f \, n^{\a\t\b}_{\s(EH)} + g^{\b\t} \nabla{^{\a}}_{\s} \f + \d^{\a}_{\s} \nabla^{\b\t} \f - \d^{(\t}_{\s} \nabla^{\b)\a} \f - g^{\a(\t} \nabla{^{\b)}}_{\s} \f \right),
\end{equation}
\begin{align}
	\hat{m}^{\a\t}_{\s(g)(5)} =& \, - \half G_5 \Box \f \, \hat{m}^{\a\t}_{\s(EH)} + \hat{G}_5 \bigg[ G^{\a\t} \nabla_{\s} \f + G{^{\a}}_{\s} \nabla^{\t} \f - G{_{\s}}^{\t} \nabla^{\a} \f - \D^{\a}_{\s\r} \nabla^{\r\t} \f  \nonumber \\
		& \, \qquad + \half \D^{\r}_{\r\s} \nabla^{\a\t} \f + \half \D^{\t}_{\r\k} \left( g^{\r\k} \nabla{^{\a}}_{\s} \f - g^{\a\k} \nabla{^{\r}}_{\s} \f + \d^{\a}_{\s} \nabla^{\r\k} \f \right) + \half \nabla{_{\s}}^{\a\t} \f  \nonumber \\
		& \, \qquad + \half \nabla{^{\t\a}}_{\s} \f - \nabla{{^{\a}}_{\s}}^{\t} \f + \D^{\r}_{\k\s} \left( \half g^{\a\k} \nabla{_{\r}}^{\t} \f - g^{\t\k} \nabla{^{\a}}_{\r} \f + \half g^{\a\t} \nabla{_{\r}}^{\k} \f \right)   \nonumber \\
		& \, \qquad + \half \d^{\a}_{\s} \left( \nabla{_{\r}}^{\r\t} \f - \nabla{^{\t\r}}_{\r} \f \right) + \d^{\t}_{\s} \left( \nabla{^{\a\r}}_{\r} \f - \nabla{_{\r}}^{\a\r} \f \right) + \half g^{\a\t} \left( \nabla{{_{\r}}^{\r}}_{\s} \f - \nabla{{_{\s}}^{\r}}_{\r} \f \right) \bigg] \nonumber \\
		& \, + \pd_{X} \hat{G}_5 \bigg[ \half \left( (\Box \f)^2 - \mbox{Tr} \Pi^2 \right) \left( \d^{\t}_{\s} \nabla^{\a} \f - \d^{\a}_{\s} \nabla^{\t} \f - g^{\a\t} \nabla_{\s} \f \right) + \Box \f \bigg( \nabla^{\t} \f \nabla{^{\a}}_{\s} \f \nonumber \\
		& \, \qquad - \nabla^{\a} \f \nabla{^{\t}}_{\s} \f + \nabla_{\s} \f \nabla^{\a\t} \f + \nabla_{\r} \f \left( \half \d^{\a}_{\s} \nabla^{\r\t} \f - \d^{\t}_{\s} \nabla^{\r\a} \f + \half g^{\a\t} \nabla{^{\r}}_{\s} \f \right) \bigg)  \nonumber \\
		& \, \qquad \left. + \half \nabla^{\a} \f \nabla^{\t\r} \f \nabla_{\r\s} \f + \half \nabla^{\a} \f \nabla_{\s\r} \f \nabla^{\r\t} \f - \nabla_{\s} \f \nabla^{\a\r} \f \nabla{_{\r}}^{\t} \f - \nabla^{\t} \f \nabla^{\a\r} \f \nabla_{\r\s} \f \right. \nonumber \\
		& \, \qquad \left. + \nabla_{\r} \f \left( \d^{\t}_{\s} \nabla^{\r\k} \f \nabla{_{\k}}^{\a} \f - \half \d^{\a}_{\s} \nabla^{\r\k} \f \nabla{_{\k}}^{\t} \f - \half g^{\a\t} \nabla^{\r\k} \f \nabla_{\k\s} \f - \half \nabla{^{\r}}_{\s} \f \nabla^{\a\t} \f \right. \right. \nonumber \\
		& \, \qquad \left. - \half \nabla^{\r\t} \f \nabla{^{\a}}_{\s} \f + \nabla^{\r\a} \f \nabla{_{\s}}^{\t} \f \right) \bigg] \nonumber \\
		& \, + \pd_{\f} \hat{G}_5 \bigg[ \half \d^{\a}_{\s} \nabla_{\r} \f \nabla^{\r\t} \f - \d^{\t}_{\s} \nabla_{\r} \f \nabla^{\r\a} \f + \half \nabla_{\s} \f \nabla^{\a\t} \f + \half g^{\a\t} \nabla_{\r} \f \nabla{^{\r}}_{\s} \f \nonumber \\
		& \, \qquad + \half \nabla^{\t} \f \nabla{^{\a}}_{\s} \f - \nabla^{\a} \f \nabla{_{\s}}^{\t} \f + \Box \f \, \left( \d^{\t}_{\s} \nabla^{\a} \f - \half \d^{\a}_{\s} \nabla^{\t} \f - \half g^{\a\t} \nabla_{\s} \f \right) \bigg],
\end{align}
\begin{align}
	\hat{u}^{\a}_{\s(g)(5)} =& \, - 2 \frac{\d \L_5}{\d g_{\a\r}} g_{\s\r} - \half G_5 \Box \f \left( \hat{u}^{\a}_{\s(EH)} + 2 \frac{\d \hat{R}}{\d g_{\a\r}} g_{\s\r} \right) \nonumber \\
& \, + \hat{G}_5 \, \bigg[ G^{\a\r} \D^{\l}_{\r\s} \nb_{\l} \f + G^{\a}_{\l} \D^{\l}_{\s\k} \nb^{\k} \f - G^{\r}_{\l} \D^{\l}_{\r\s} \nb^{\a} \f + \half \D^{\a}_{\s\r} \left( \nb{_{\l}}^{\l\r} \f - \nb{^{\r\l}}_{\l} \f \right) \nonumber \\
& \, \quad + \D^{\r}_{\r\s} \left( \nb{^{\a\l}}_{\l} \f - \nb{_{\l}}^{\l\a} \f \right) + \half \D^{\r}_{\s\l} \left( \nb{_{\r}}^{\a\l} \f + \nb{^{\l\a}}_{\r} \f  - 2 \nb{^{\a\l}}_{\r} \f \right. \nonumber \\
& \, \quad \left. + g^{\a\l} \left( \nb{^{\k}}_{\k\r} \f - \nb{_{\r\k}}^{\k} \f \right) \right) - \half \nb_{\r} \D^{\a}_{\s\l} \nb^{\r\l} \f + \nb_{\r} \D^{\l}_{\l\s} \nb^{\a\r} \f - \half \nb_{\r} \D^{\r}_{\s\l} \nb^{\a\l} \f \nonumber \\
& \, \quad - \half \nb^{\r} \D^{\l}_{\s\r} \nb{_{\l}}^{\a} \f + \nb^{\a} \D^{\l}_{\s\r} \nb{_{\l}}^{\r} \f - \half \nb^{\r} \D^{\l}_{\s\k} \nb_{\l\r} \f \, g^{\a\k} - \half \D^{\a}_{\r\l} \D^{\r}_{\s\k} \nb^{\k\l} \f \nonumber \\
& \, \quad - \half \D^{\r}_{\l\k} \D^{\l}_{\s\r} \nb^{\a\k} \f + \D^{\r}_{\r\l} \D^{\l}_{\s\k} \nb^{\a\k} \f - \half \D^{\r}_{\l\k} \D^{\l}_{\s\iota} \nb{_{\r}}^{\a} \f \, g^{\k\iota} - \half \D^{\r}_{\l\k} \D^{\l}_{\s\iota} \nb{_{\r}}^{\iota} \f \, g^{\a\k} \nonumber \\
& \, \quad + \D^{\r}_{\l\k} \D^{\l}_{\s\iota} \nb{_{\r}}^{\k} \f \, g^{\a\iota} - \frac{3}{4} \bar{R}_{\s\l} \nb^{\a\l} \f - \frac{3}{4} \bar{R}{^{\r}}_{\s\l\k} \nb{_{\r}}^{\l} \f \, g^{\a\k} \bigg] \nonumber \\
& \, + \pd_X \hat{G}_5 \, \bigg[ \left( \D^{[\r}_{\r\s} \nb^{\a]} \f - \half \D^{\r}_{\s\l} \nb_{\r} \f \, g^{\a\l} \right) \left( \left( \Box \f \right)^2 - \mbox{Tr} \Pi^2 \right) + \Box \f \left( \half \D^{\a}_{\s\r} \nb_{\l} \f \nb^{\r\l} \f \right. \nonumber \\
& \, \quad - \D^{\r}_{\r\s} \nb_{\l} \f \nb^{\a\l} \f + \D^{\r}_{\s\l} \Big( \nb_{\r} \f \nb^{\a\l} \f - \nb^{\a} \f \nb{_{\r}}^{\l} \f + \nb^{\l} \f \nb{_{\r}}^{\a} \f  \nonumber \\
& \, \quad  \left. + \half \nb_{\k} \f \nb{_{\r}}^{\k} \f \, g^{\a\l} \Big) \right) - \half \D^{\a}_{\s\r} \nb_{\l} \f \nb{_{\k}}^{\r} \f \nb^{\l\k} \f + \D^{\r}_{\r\s} \nb_{\l} \f \nb{_{\k}}^{\a} \f \nb^{\l\k} \f \nonumber \\
& \, \quad + \D^{\r}_{\s\l} \bigg( \nb_{\k} \f \left( \nb{_{\r}}^{\l} \f \nb^{\a\k} \f - \half \nb{_{\r}}^{\a} \f \nb^{\l\k} \f - \half \nb{_{\r}}^{\k} \f \nb^{\a\l} \f - \half \nb{_{\iota}}^{\k} \f \nb{_{\r}}^{\iota} \f \, g^{\a\l} \right) \nonumber \\
& \, \quad - \nb_{\r} \f \nb{_{\k}}^{\a} \f \nb^{\l\k} \f + \nb^{\a} \f \nb{_{\k}}^{\l} \f \nb{_{\r}}^{\k} \f - \nb^{\l} \f \nb{_{\k}}^{\a} \f \nb{_{\r}}^{\k} \f \bigg) \bigg] \nonumber \\
& \, + {\pd}_{\f}{\hat{G}_5}\, \bigg[
   \frac{1}{2}\, {\D}^{\a}_{\s \r} {\nb}_{\l}{\f}\,  {\nb}^{\r \l}{\f}\, 
 - {\D}^{\r}_{\s \r} {\nb}_{\l}{\f}\,  {\nb}^{\a \l}{\f}\, 
 + \Box \f \Big( - \frac{1}{2}\, {\D}^{\a}_{\s \r} {\nb}^{\r}{\f}  \nonumber \\
& \, \quad
 - \frac{1}{2}\, {\D}^{\r}_{\s \l} {\nb}_{\r}{\f}\,  {g}^{\a \l}
 +  {\D}^{\r}_{\s \r} {\nb}^{\a}{\f} \Big)
 + \D^{\r}_{\s\l} \Big( \frac{1}{2}\, {\nb}^{\l}{\f}\,  {\nb}{_{\r}}^{\a}{\f}
 - {\nb}^{\a}{\f}\,  {\nb}{_{\r}}^{\l}{\f} \nonumber \\
& \, \quad
 + \frac{1}{2}\, {\nb}_{\k}{\f}\,  {\nb}{_{\r}}^{\k}{\f}\,  {g}^{\a \l} 
 + \frac{1}{2}\, {\nb}_{\r}{\f}\,  {\nb}^{\a \l}{\f} \Big) \bigg].
\end{align}

\section{Superpotential $\hat{i}^{\a\b}$}

The general formula for the superpotential reads [see Eq. (55) in Ref. \citen{petrovmain}]
\begin{equation}
	\hat{i}^{\a\b} = \left( \frac{2}{3} \bn_{\l} \hat{n}^{[\a\b]\l}_{\s} - \hat{m}^{[\a\b]}_{\s} \right) \x^{\s} - \frac{4}{3} \hat{n}^{[\a\b]\l}_{\s} \bn_{\l} \x^{\s}. \label{superpot}
\end{equation}
The conserved current $\hat{i}^{\a}$, given by (\ref{icur}), is generated by the superpotential as a divergence: $\hat{i}^{\a} = \pd_{\b} \hat{i}^{\a\b}$.

Since we have two fields, we split the superpotential for the scalar and for the tensor field and for each Lagrangian, as in the case of the conserved currents. We write:
\begin{equation}
	\hat{i}^{\a\b} = \sum_{i=2}^5 \left( \hat{i}^{\a\b}_{(\f)(i)} + \hat{i}^{\a\b}_{(g)(i)} \right).
\end{equation}
The coefficients $\hat{n}^{\a\t\b}_{\s}$ vanish for the scalar field, see (\ref{fcurn}), so the superpotential for $\varphi$-field reduces to 
\begin{equation}
	\hat{i}^{\a\b}_{(\f)} = - \hat{m}^{[\a\b]}_{\s(\f)} \x^{\s}. \label{supotscalar}
\end{equation}
Moreover, all coefficients $\hat{m}^{\a\b}_{\s}$ of scalar parts are symmetrical in $(\a,\b)$ [c.f. (\ref{fcurm})], hence $\hat{i}^{\a\b}_{(\f)(i)} = 0$ for all the Lagrangians. We thus need to calculate only the metric field part of the superpotential. For Lagrangian $\Lh_2$, the coefficients $\hat{n}^{\a\t\b}_{\s(g)(2)}$ and $\hat{m}^{\a\t}_{\s(g)(2)}$ vanish, see Sec. \ref{coefl2}, and so does the superpotential $\hat{i}^{\a\b}_{(g)(2)}$.

Regarding the Lagrangian $\Lh_3$, as the coefficient $\hat{n}^{\a\t\b}_{\s(g)(3)}$ is also vanishing, we get the similar situation as for the scalar field and the superpotential formula reduces to the same form as in (\ref{supotscalar}). Then, plugging (\ref{l3gm}) into (\ref{supotscalar}), we get 
\begin{equation}
	\hat{i}^{\a\b}_{(3)} = \hat{i}^{\a\b}_{(g)(3)} = 2 \hat{G}_3 \d^{[\a}_{\s} \nb^{\b]} \varphi \, \xi^{\s}. \label{iab3}
\end{equation}

For the Lagrangian $\Lh_4$ we split the superpotential into two parts -- we will explicitly exclude the part originating from the Einstein-Hilbert part of current coefficients $\hat{m}^{\a\t}_{\s}$ and $\hat{n}^{\a\t\b}_{\s}$. Generally speaking, if the structure of coefficients $\hat{m}^{\a\t}_{\s}$ and $\hat{n}^{\a\t\b}_{\s}$ is
\begin{equation}
	\hat{m}^{\a\t}_{\s} = F \, \hat{m}^{\a\t}_{\s(EH)} + \hat{m}^{\a\t}_{\s(rest)}, \qquad \hat{n}^{\a\t\b}_{\s} = F \, \hat{n}^{\a\t\b}_{\s(EH)} + \hat{n}^{\a\t\b}_{\s(rest)}, \label{supcoefsplit}
\end{equation}
where $F$ is an arbitrary function, we obtain, after plugging (\ref{supcoefsplit}) into (\ref{superpot}), the following splitting of the superpotential:
\begin{equation}
	\hat{i}^{\a\b} = F \, \hat{i}^{\a\b}_{(EH)} + \hat{i}^{\a\b}_{(rest)}, \label{supsplit}
\end{equation}
with the latter part given by the expression
\begin{equation}
	\hat{i}^{\a\b}_{(rest)} = \left[ \frac{2}{3} \left( \bn_{\l} F \, \hat{n}^{[\a\b]\l}_{\s(EH)} + \D^{\k}_{\k\l} \hat{n}^{[\a\b]\l}_{\s(rest)} + \sqrt{-g} \, \bn_{\l} n^{[\a\b]\l}_{\s(rest)} \right) - \hat{m}^{[\a\b]}_{\s(rest)} \right] \x^{\s} - \frac{4}{3} \hat{n}^{[\a\b]\l}_{\s(rest)} \bn_{\l} \x^{\s}. 
\end{equation}
Then, following this scheme for the Lagrangian $\Lh_4$ with $F = G_4$, we get
\begin{align}
	\hat{i}^{\a\b}_{(4)} &= G_4 \, \hat{i}^{\a\b}_{(EH)} + \hat{i}^{\a\b}_{(4)(rest)},  \label{iab4} \\
	\hat{i}^{\a\b}_{(4)(rest)} &= 4 \left[ \pd_X \hat{G}_4 \left( \d^{[\a}_{\s} \nb^{\b]\r} \f \nb_{\r} \f - \d^{[\a}_{\s} \nb^{\b]} \f \, \Box \f - \nb^{[\a} \f \nb{^{\b]}}_{\s} \f \right) - \pd_{\f} \hat{G}_4 \, \d^{[\a}_{\s} \nb^{\b]} \f \right] \x^{\s}. \label{iab4rest}
\end{align}

Finally, for the last Lagrangian $\Lh_5$ we have
\begin{equation}
	\hat{i}^{\a\b}_{(5)} = - \half G_5 \, \Box \f \, \hat{i}^{\a\b}_{(EH)} + \hat{i}^{\a\b}_{(5)(rest)}. \label{iab5}
\end{equation}
The calculation of $\hat{i}^{\a\b}_{(5)(rest)}$ is done similarly as for the coefficients $\hat{m}^{\a\t}_{\s}$ and $\hat{u}^{\a}_{\s}$ for Lagrangians $\Lh_4$ and $\Lh_5$. The result turns out to be 
\begin{align}
	\hat{i}^{\a\b}_{(5)(rest)} =& \, \Big[ \hat{G}_5 \Big( 2 \d^{[\a}_{\s} \nb{^{\b]\l}}_{\l} \f + 2 \nb{^{[\a\b]}}_{\s} \f - 2 \d^{[\a}_{\s} \nb{_{\l}}^{\b]\l} \f - \D^{\r}_{\s\k} g^{\k[\a} \nb{^{\b]}}_{\r} \f \nonumber \\
		& \, \qquad  + \D^{[\a}_{\s\r} \nb^{\b]\r} \f - 2 G^{[\a}_{\s} \nb^{\b]} \f \Big) \nonumber \\
		& \, + 2 \, \pd_{\f} \hat{G}_5 \left( \d^{[\a}_{\s} \nb^{\b]} \f \, \Box \f + \nb^{[\a} \f \nb{^{\b]}}_{\s} \f - \d^{[\a}_{\s} \nb^{\b]\r} \f \nb_{\r} \f \right) \nonumber \\
		& \, + \pd_X \hat{G}_5 \left( -2 \d^{[\a}_{\s} \nb^{\b]\r} \f \nb_{\r} \f \, \Box \f + 2 \nb_{\r} \f \nb{_{\s}}^{[\a} \f \nb^{\b]\r} \f + 2 \d^{[\a}_{\s} \nb^{\b]\r} \f \nb_{\r\k} \f \nb^{\k} \f \right. \nonumber \\
		& \, \qquad + \d^{[\a}_{\s} \nb^{\b]} \f (\Box \f)^2 - \d^{[\a}_{\s} \nb^{\b]} \f \, \mbox{Tr} \, \Pi^2 + 2 \nb^{[\a} \f \nb{^{\b]}}_{\s} \f \, \Box \f \nonumber \\
		& \, \qquad \left. - 2 \nb^{[\a} \f \nb^{\b]\r} \f \nb_{\r\s} \f \right) \Big] \x^{\s} + \hat{G}_5 \left[ \d^{[\a}_{\s} \nb^{\b]\l} \f - g^{\l[\a} \nb{^{\b]}}_{\s} \f \right] \bn_{\l} \x^{\s}. \label{iab5rest}
\end{align}

\section{Superpotentials associated with nonlinear and linear perturbations}

Superpotentials associated with the background are obtained simply by replacing all $g_{\m\n}$ and $\varphi$ by $\bar{g}_{\m\n}$ and $\bar{\f}$ and, consequently, all covariant derivatives $\nb$ are replaced by ones with respect to the background metric $\bn$, and the connections difference $\Delta^{\l}_{\m\n}$ vanishes. The superpotentials can be made relevant for (possibly large) perturbations if we consider the difference between the ``total'' and the background superpotentials as
follows:
\begin{equation}
	\hat{I}^{\a\b} = \hat{i}^{\a\b} - \bar{\hat{i}}^{\a\b}.
\end{equation}
If the quantities $\varphi$, $g_{\m\n}$ and $\bar{\varphi}$, $\bar{g}_{\m\n}$ are solutions of the field equations for both physical and background spacetimes, we can construct relative superpotentials and associated conserved charges for specific physical problems.

In this section, indices are raised and lowered with the metric $\bar{g}_{\m\n}$ only. For the Lagrangian $\Lh_3$ we have the background superpotential
\begin{equation}
	\bar{\hat{i}}^{\a\b}_{(3)} = 2 \bar{\hat{G}}_3 \d^{[\a}_{\s} \bn^{\b]} \bar{\f} \, \xi^{\s},
\end{equation}
where the notation $\bar{\hat{F}}$ means $\sqrt{-\bar{g}} F(\bar{\f}, \bar{X})$ with $F$ being arbitrary function of $\f$ and $X$; naturally, $\bar{X} = \frac{1}{2} \bar{g}^{\m\n} \bn_{\m}{\bar{\f}}\bn_{\n}{\bar{\f}}$. For the Lagrangian $\Lh_4$, we have the splitting (\ref{iab4}), hence
\begin{equation}
	\bar{\hat{i}}^{\a\b}_{(4)} = \bar{G}_4 \bar{\hat{i}}^{\a\b}_{(EH)} + \bar{\hat{i}}^{\a\b}_{(4)(rest)},
\end{equation}
where the second part of the expression is given by
\begin{equation}
	\bar{\hat{i}}^{\a\b}_{(4)(rest)} = 4 \left[ \pd_X \bar{\hat{G}}_4 \left( \d^{[\a}_{\s} \bn^{\b]\r} \bar{\f} \bn_{\r} \bar{\f} - \d^{[\a}_{\s} \bn^{\b]} \bar{\f} \, \bar{\Box} \bar{\f} - \bn^{[\a} \bar{\f} \bn{^{\b]}}_{\s} \bar{\f} \right) - \pd_{\f} \bar{\hat{G}}_4 \, \d^{[\a}_{\s} \bn^{\b]} \bar{\f} \right] \x^{\s};
\end{equation}
the obvious notation $\bar{\Box} = \bar{g}^{\m\n} \bn_{\m} \bn_{\n}$ was used. For the last Lagrangian $\Lh_5$, we use the splitting (\ref{iab5}), consequently, we have the expression for the background fields as follows:
\begin{equation}
	\bar{\hat{i}}^{\a\b}_{(5)} = - \half \bar{G}_5 \, \bar{\Box} \bar{\f} \, \bar{\hat{i}}^{\a\b}_{(EH)} + \bar{\hat{i}}^{\a\b}_{(5)(rest)},	
\end{equation}
with the following lengthy expression
\begin{align}
	\bar{\hat{i}}^{\a\b}_{(5)(rest)} =& \, \Big[ \bar{\hat{G}}_5 \left( 2 \d^{[\a}_{\s} \bn{^{\b]\l}}_{\l} \bar{\f} + 2 \bn{^{[\a\b]}}_{\s} \bar{\f} - 2 \d^{[\a}_{\s} \bn{_{\l}}^{\b]\l} \bar{\f} - 2 \bar{G}^{[\a}_{\s} \bn^{\b]} \bar{\f} \right) \nonumber \\
		& \, + 2 \, \pd_{\f} \bar{\hat{G}}_5 \left( \d^{[\a}_{\s} \bn^{\b]} \bar{\f} \, \bar{\Box} \bar{\f} + \bn^{[\a} \bar{\f} \bn{^{\b]}}_{\s} \bar{\f} - \d^{[\a}_{\s} \bn^{\b]\r} \bar{\f} \bn_{\r} \bar{\f} \right) \nonumber \\
		& \, + \pd_X \bar{\hat{G}}_5 \left( -2 \d^{[\a}_{\s} \bn^{\b]\r} \bar{\f} \bn_{\r} \bar{\f} \, \bar{\Box} \bar{\f} + 2 \bn_{\r} \bar{\f} \bn{_{\s}}^{[\a} \bar{\f} \bn^{\b]\r} \bar{\f} + 2 \d^{[\a}_{\s} \bn^{\b]\r} \bar{\f} \bn_{\r\k} \bar{\f} \bn^{\k} \bar{\f} \right. \nonumber \\
		& \, \qquad + \d^{[\a}_{\s} \bn^{\b]} \bar{\f} (\bar{\Box} \bar{\f})^2 - \d^{[\a}_{\s} \bn^{\b]} \bar{\f} \, \mbox{Tr} \, \bar{\Pi}^2 + 2 \bn^{[\a} \bar{\f} \bn{^{\b]}}_{\s} \bar{\f} \, \bar{\Box} \bar{\f} \nonumber \\
		& \, \qquad \left. - 2 \bn^{[\a} \bar{\f} \bn^{\b]\r} \bar{\f} \bn_{\r\s} \bar{\f} \right) \Big] \x^{\s} + \bar{\hat{G}}_5 \left[ \d^{[\a}_{\s} \bn^{\b]\l} \bar{\f} - g^{\l[\a} \bn{^{\b]}}_{\s} \bar{\f} \right] \bn_{\l} \x^{\s} \\
		=& \, \left[ \bar{\hat{G}}_5 \, \bar{i}^{\a\b}_{\s (G_5)} + \pd_{\f} \bar{\hat{G}}_5 \, \bar{i}^{\a\b}_{\s (\pd_{\f} G_5)} + \pd_X \bar{\hat{G}}_5 \, \bar{i}^{\a\b}_{\s (\pd_X G_5)} \right] \x^{\s} + \bar{\hat{G}}_5 \, \bar{i}^{\a\b\l}_{\s (\nb G_5)} \bn_{\l} \x^{\s},
\end{align}
where the identity $\bar{\D}^{\l}_{\m\n} = 0$ and the notation $\mbox{Tr} \, \bar{\Pi}^2 = \bn_{\m\n} \bar{\f} \bn^{\m\n} \bar{\f}$ were used; we also introduced notation $\bar{i}^{\a\b}_{\s (G_5)}$, $\bar{i}^{\a\b}_{\s (\pd_{\f} G_5)}$, $\bar{i}^{\a\b}_{\s (\pd_X G_5)}$ and $\bar{i}^{\a\b\l}_{\s (\nb G_5)}$, denoting the terms in brackets appearing at various derivatives of function $G_5$. These will be subsequently used in the expressions for the linearized superpotential.

The linearization of these superpotentials is done by assuming the metric and the scalar field in the form $g_{\m\n} = \bar{g}_{\m\n} + \varepsilon h_{\m\n}$ and $\f = \bar{\f} + \varepsilon \d \f$ in superpotentials (\ref{iab3}), (\ref{iab4})-(\ref{iab5rest}) and keeping terms only of the first order in $\varepsilon$. We also have $g^{\m\n} = \bar{g}^{\m\n} - \varepsilon h^{\m\n} + O(\varepsilon^2)$, with $h^{\m\n} = \bar{g}^{\m\r} \bar{g}^{\n\s} h_{\r\s}$. Typically, for a term $H$ the quantity $\d H$ means: $H = \bar{H} + \varepsilon \d H + O(\varepsilon^2)$, where in $\bar{H}$ every quantity was replaced by its background counterpart. The linearization of the following expressions is obtained easily:
\begin{align}
    G_i(\f, X) &= G_i(\bar{\f}, \bar{X}) + \varepsilon \left( \pd_{\f} G_i(\bar{\f}, \bar{X}) \d \f + \pd_{X} G_i(\bar{\f}, \bar{X}) \d X \right) + O(\varepsilon^2), \\
	\d X &= - \bn^{\m} \bar{\f} \bn_{\m} \d \f - \frac{1}{2} h^{\m\n} \bn_{\m} \bar{\f} \bn_{\n} \bar{\f}, \\
	\d \Delta^{\l}_{\m\n} &= \frac{1}{2} \bar{g}^{\l\k} \left( \bn_{\m} h_{\n\k} + \bn_{\n} h_{\m\k} - \bn_{\k} h_{\m\n} \right), \label{dChdifflin} \\
	\nb_{\m\n} \f &= \bn_{\m\n} \bar{\f} + \varepsilon \left( \bn_{\m\n} \d \f - \d \D^{\r}_{\m\n} \bn_{\r} \bar{\f} \right) + O(\varepsilon^2), \\
	\sqrt{-g} &= \sqrt{-\bar{g}} \left( 1 + \frac{1}{2} \varepsilon h \right) + O(\varepsilon^2),
\end{align}
where $h = h^{\m}_{\m} = \bar{g}^{\m\n} h_{\m\n}$. 

For the superpotentials we have $\hat{i}^{\a\b}_{(i)} = \bar{\hat{i}}^{\a\b}_{(i)} + \varepsilon \, \d \hat{i}^{\a\b}_{(i)} + O(\varepsilon^2)$, and for the Lagrangian $\Lh_3$, we obtain
\begin{align}
	\d \hat{i}^{\a\b}_{(3)} &= 2 \d^{[\a}_{\s} \left[ \bar{\hat{G}}_3 \left( \bn^{\b]} \d \f + \half h \bn^{\b]} \bar{\f} - h^{\b]\r} \bn_{\r} \bar{\f} \right) + \pd_{\f} \bar{\hat{G}}_3 \, \d \f \, \bn^{\b]} \bar{\f} \, \right. \nonumber \\
	& \qquad \qquad \left. + \pd_X \bar{\hat{G}}_3 \bn^{\b]} \bar{\f} \left( \half \d g^{\m\n} \bn_{\m} \bar{\f} \bn_{\n} \bar{\f} - \bn_{\r} \bar{\f} \bn^{\r} \d \f \right) \right] \x^{\s}.
\end{align}
Assuming $\bar{\f} = \mbox{const.}$ we get simply $\d \hat{i}^{\a\b}_{(3)} = 2 \bar{\hat{G}}_3 \d^{[\a}_{\s}  \bn^{\b]} \d \f \, \x^{\s} $.

The splitting of the superpotential (\ref{supsplit}) leads to the following decomposition:
\begin{equation}
	\d \hat{i}^{\a\b} = \d F \, \bar{\hat{i}}^{\a\b}_{(EH)} + \bar{F} \, \d \hat{i}^{\a\b}_{(EH)} + \d \hat{i}^{\a\b}_{(rest)},
\end{equation}
with $\bar{F}$ simply denoting $F(\bar{\f}, \bar{X})$ and $\d F = \pd_{\f} \bar{F} \, \d \f + \pd_{X} \bar{F} \, \d X$. In the case of the Lagrangian $\Lh_4$, the decomposition looks as follows:
\begin{equation}
	\d \hat{i}^{\a\b}_{(4)} = \d G_4 \bar{\hat{i}}^{\a\b}_{(EH)} + \bar{G}_4 \d \hat{i}^{\a\b}_{(EH)} + \d \hat{i}^{\a\b}_{(4)(rest)},
\end{equation}
and for the $\d \hat{i}^{\a\b}_{(4)(rest)}$ we arrive at the expression:
\begin{align}
\d \hat{i}^{\a\b}_{(4)(rest)} =& \, 4 \bigg[ \pd_{\f} \bar{\hat{G}}_4 \, \d^{[\a}_{\s} \left( h^{\b]\r} \bn_{\r} \bar{\f} - \bn^{\b]} \d \f - \half h \bn^{\b]} \bar{\f} \right) - \pd_{\f\f} \bar{\hat{G}}_4 \, \d \f \, \d^{[\a}_{\s} \bn^{\b]} \bar{\f}  \nonumber \\
	& \, + \pd_X \bar{\hat{G}}_4 \Big[ \d^{[\a}_{\s} \left( \bn^{\b]\r} \bar{\f} \bn_{\r} \d \f + \bn^{\b]\r} \d \f \bn_{\r} \bar{\f} - \bn^{\b]} \bar{\f} \, \bar{\Box} \d \f - \bn^{\b]} \d \f \, \bar{\Box} \bar{\f} \right) \nonumber \\
	& \, \qquad - \bn^{[\a} \bar{\f} \bn{^{\b]}}_{\s} \d \f - \bn^{[\a} \d \f \bn{^{\b]}}_{\s} \bar{\f} + \half h \big( \d^{[\a}_{\s} \bn^{\b]\r} \bar{\f} \bn_{\r} \bar{\f} - \d^{[\a}_{\s} \bn^{\b]} \bar{\f} \, \bar{\Box} \bar{\f} \nonumber \\
	& \, \qquad  - \bn^{[\a} \bar{\f} \bn{^{\b]}}_{\s} \bar{\f} \big) - \d^{[\a}_{\s} \bar{g}^{\b]\r} \bn_{\k} \bar{\f} \bn^{\l} \bar{\f} \, \d \D^{\k}_{\r\l} + \d^{[\a}_{\s} \bn^{\b]} \bar{\f} \bn_{\r} \bar{\f} \, \d \D^{\r}_{\k\l} \bar{g}^{\k\l}  \nonumber \\
	& \, \qquad + \bn_{\r} \bar{\f} \bn^{[\a} \bar{\f} \, \bar{g}^{\b]\k} \, \d \D^{\r}_{\k\s} + \d^{[\a}_{\s} \big( - \bn{^{\b]}}_{\r} \bar{\f} \bn_{\k} \bar{\f} \, h^{\r\k} - h^{\b]\r} \bn^{\k} \bar{\f} \bn_{\r\k} \bar{\f}   \nonumber \\
	& \, \qquad + \bn^{\b]} \bar{\f} \bn_{\r\k} \bar{\f} \, h^{\r\k} + h^{\b]\r} \bn_{\r} \bar{\f} \, \bar{\Box} \bar{\f} \big) + \bn^{[\a} \bar{\f} \, h^{\b]\r} \bn_{\r\s} \bar{\f} + h^{\r[\a} \bn{^{\b]}}_{\s} \bar{\f} \bn_{\r} \bar{\f} \Big] \nonumber \\
	& \, + \pd_{X\f} \bar{\hat{G}}_4 \Big[ \d \f \left( \d^{[\a}_{\s} \bn^{\b]\r} \bar{\f} \bn_{\r} \bar{\f} - \d^{[\a}_{\s} \bn^{\b]} \bar{\f} \, \bar{\Box} \bar{\f} - \bn^{[\a} \bar{\f} \bn{^{\b]}}_{\s} \bar{\f} \right)  - \d X \, \d^{[\a}_{\s} \bn^{\b]} \bar{\f}  \Big] \nonumber \\
	& \, + \pd_{XX} \bar{\hat{G}}_4 \, \d X \left( \d^{[\a}_{\s} \bn^{\b]\r} \bar{\f} \bn_{\r} \bar{\f} - \d^{[\a}_{\s} \bn^{\b]} \bar{\f} \, \bar{\Box} \bar{\f} - \bn^{[\a} \bar{\f} \bn{^{\b]}}_{\s} \bar{\f} \right) \bigg] \x^{\s}.
\end{align}
After setting $\bar{\f} = \mbox{const.}$, we obtain $\D \hat{i}^{\a\b}_{(4)(rest)} = - \pd_{\f} \bar{\hat{G}}_4 \d^{[\a}_{\s} \bn^{\b]} \d \f \, \x^{\s}$.

In the superpotential for the Lagrangian $\Lh_5$ (\ref{iab5rest}), we observe two terms for which the linearization is not so obvious. It is the Einstein tensor $G_{\m\n}$ and the third derivatives of scalar field $\nb_{\a\b\g} \f$. Let us examine them closer. Using (\ref{covriemann}) and realizing that $\D^{\l}_{\m\n}$ is already of the first order, we obtain the Riemann tensor and, by contraction, the linearized Ricci tensor
\begin{align}
	R{^{\l}}_{\t\r\s} &= \bar{R}{^{\l}}_{\t\r\s} + \varepsilon \left( \bn_{\r} \d \D^{\l}_{\t\s} - \bn_{\s} \d \D^{\l}_{\t\r} \right) + O(\varepsilon^2), \nonumber \\
	R_{\t\s} &= \bar{R}_{\t\s} + \varepsilon \left( \bn_{\l} \d \D^{\l}_{\t\s} - \bn_{\s} \d \D^{\l}_{\l\t} \right) + O(\varepsilon^2). \label{Riemannlin}
\end{align}
The Ricci scalar $R = g^{\t\s} R_{\t\s}$ is then linearized as follows
\begin{equation}
	R = \bar{R} + \varepsilon \left( - h^{\t\s} \bar{R}_{\t\s} + \bar{g}^{\t\s} \left( \bn_{\l} \d \D^{\l}_{\t\s} - \bn_{\s} \d \D^{\l}_{\l\t} \right) \right) + O(\varepsilon^2),
\end{equation}
leading to the final expression for the linearized Einstein tensor $G_{\m\n} = \bar{G}_{\m\n} + \varepsilon \, \d G_{\m\n} + O(\varepsilon^2)$:
\begin{equation}
	\d G_{\m\n} = \bn_{\l} \d \D^{\l}_{\m\n} - \bn_{\m} \d \D^{\l}_{\l\n} - \half h_{\m\n} \bar{R} - \half \bar{g}_{\m\n} \left( - h^{\t\s} \bar{R}_{\t\s} + \bar{g}^{\t\s} \left( \bn_{\l} \d \D^{\l}_{\t\s} - \bn_{\s} \d \D^{\l}_{\l\t} \right) \right). \label{dGmnlin}
\end{equation}

The linearization of the third covariant derivative of the scalar field proceeds as follows. Converting the outermost derivative into a background one, we obtain
\begin{equation}
	\nb_{\a\b\g} \f = \bn_{\a}\left( \nb_{\b\g} \f \right) - \D^{\r}_{\a\b} \nb_{\g\r} \f - \D^{\r}_{\g\a} \nb_{\b\r} \f,
\end{equation}
and after substituting from (\ref{d2fcov}), we have
\begin{equation}
	\nb_{\a\b\g} \f = \bn_{\a\b\g} \f - 3 \D^{\r}_{(\a\b} \bn_{\g)\r} \f - \left( \bn_{\a} \D^{\r}_{\b\g} - \D^{\k}_{\a\b} \D^{\r}_{\g\k} - \D^{\k}_{\g\a} \D^{\r}_{\b\k} \right) \bn_{\r} \f. \label{d3fback}
\end{equation}
This is then linearized in a straightforward way into
\begin{equation}
	\nb_{\a\b\g} \f = \bn_{\a\b\g} \bar{\f} + \varepsilon \left( \bn_{\a\b\g} \d \f - 3 \d \D^{\r}_{(\a\b} \bn_{\g)\r} \bar{\f} - \bn_{\a} \d \D^{\r}_{\b\g} \bn_{\r} \bar{\f} \right) + O(\varepsilon^2). \label{d3flin}
\end{equation}
Hence, we can write
\begin{equation}
	\d\left( \nb_{\a\b\g} \f \right) = \bn_{\a\b\g} \d \f - 3 \d \D^{\r}_{(\a\b} \bn_{\g)\r} \bar{\f} - \bn_{\a} \d \D^{\r}_{\b\g} \bn_{\r} \bar{\f}. \label{d3flinfin}
\end{equation}
Considering antisymmetrization of two outermost derivatives in (\ref{d3fback}), we obtain
\begin{equation}
	\nb_{[\a\b]\g} \f = \bn_{[\a\b]\g} \f - \left( \bn_{[\a} \D^{\r}_{\b]\g} - \D^{\k}_{\g[\a} \D^{\r}_{\b]\k} \right) \bn_{\r} \f,
\end{equation}
and rewriting the terms in brackets using (\ref{covriemann}) we arrive at the familiar result
\begin{equation}
	2 \nb_{[\a\b]\g} \f = 2 \bn_{[\a\b]\g} \f - \left( R{^{\r}}_{\g\a\b} - \bar{R}{^{\r}}_{\g\a\b} \right) \bn_{\r} \f.
\end{equation}
Linearizing the last expression using (\ref{Riemannlin}) gives
\begin{equation}
	2 \nb_{[\a\b]\g} \f = 2 \bn_{[\a\b]\g} \bar{\f} + \varepsilon \left( 2 \bn_{[\a\b]\g} \d \f - \left( \bn_{\a} \d \D^{\r}_{\g\b} - \bn_{\b} \d \D^{\r}_{\g\a} \right) \bn_{\r} \bar{\f} \right) + O(\varepsilon^2),
\end{equation}
which is in agreement with result (\ref{d3flin}). The substitution $2 \bn_{[\a\b]\g} \d \f = - \bar{R}{^{\r}}_{\g\a\b} \bn_{\r} \d \f$ completes the result, so that we have
\begin{equation}
	\d\left( \nb_{[\a\b]\g} \f \right) = - \half \bar{R}{^{\r}}_{\g\a\b} \bn_{\r} \d \f - \bn_{[\a} \d \D^{\r}_{\b]\g} \bn_{\r} \bar{\f}.
\end{equation}

The last superpotential has the following form
\begin{equation}
	\d \hat{i}^{\a\b}_{(5)} = - \half \d \left( G_5 \Box \f \right) \bar{\hat{i}}^{\a\b}_{(EH)} - \half \bar{G}_5 \bar{\Box} \bar{\f} \, \d \hat{i}^{\a\b}_{(EH)} + \d \hat{i}^{\a\b}_{(5)(rest)},
\end{equation}
with
\begin{equation}
	\d \hat{i}^{\a\b}_{(5)(rest)} = \left[ \hat{Q}^{\a\b}_{\s} + \hat{R}^{\a\b}_{\s} + \hat{S}^{\a\b}_{\s} \right] \x^{\s} + \hat{P}^{\a\b\l}_{\s} \bn_{\l} \x^{\s},
\end{equation}
where the terms above are given by the following lengthy expressions:
\begin{align}
	\hat{P}^{\a\b\l}_{\s} = & \, \bar{\hat{G}}_5 \Big[ \d^{[\a}_{\s} \left( - h^{\b]\r} \bn{_{\r}}^{\l} \bar{\f} - \bn{^{\b]}}_{\t} \bar{\f} \, h^{\t\l} + \bn^{\b]\l} \d \f - \bn_{\r} \bar{\f} \, \bar{g}^{\b]\k} \d \D^{\r}_{\k\t} \bar{g}^{\t\l} \right) \nonumber \\
		& \, \qquad - \bar{g}^{\l[\a} \left( - h^{\b]\r} \bn_{\r\s} \bar{\f} + \bn{^{\b]}}_{\s} \d \f - \bn_{\t} \bar{\f} \, \bar{g}^{\b]\k} \d \D^{\t}_{\k\s} \right) + h^{\l[\a} \bn^{^{\b]}}_{\s} \bar{\f} \Big] \nonumber \\
		& \, + \left( \pd_{\f} \bar{\hat{G}}_5 \d \f + \pd_X \bar{\hat{G}}_5 \d X + \half \bar{\hat{G}}_5 \, h \right) \bar{i}^{\a\b\l}_{\s (\nb G_5)},
\end{align}
\begin{align}
	\hat{Q}^{\a\b}_{\s} = & \, \bar{\hat{G}}_5 \Big[ 2 \d^{[\a}_{\s} \Big( \hspace{-0.3em} - \hspace{-0.2em} h^{\b]\r} \bn{_{\r\l}}^{\l} \bar{\f} - \bn{^{\b]}}_{\k\l} \bar{\f} \, h^{\k\l} + \bar{g}^{\b]\r} \bar{g}^{\k\l} \d(\nb_{\r\k\l} \f) + h^{\b]\r} \bn{_{\l\r}}^{\l} \bar{\f} + \bn{{_{\k}}^{\b]}}_{\l} \bar{\f} \, h^{\k\l} \nonumber \\
		& \, \qquad - \bar{g}^{\b]\r} \bar{g}^{\k\l} \d( \nb_{\l\r\k} \f ) \Big) - 2 h^{\r[\a} \bn{{_{\r}}^{\b]}}_{\s} \bar{\f} - 2 h^{\k[\b} \bn{^{\a]}}_{\k\s} \bar{\f} + 2 \bar{g}^{\r[\a} \bar{g}^{\b]\k} \d ( \nb_{\r\k\s} \f ) \nonumber \\
		& \, \qquad - \d \D^{\r}_{\s\k} \bar{g}^{\k[\a} \bn{^{\b]}}_{\r} \bar{\f} + \d \D^{[\a}_{\s\r} \bn^{\b]\r} \bar{\f}  - 2 \d G_{\s\r} \bar{g}^{\r[\a} \bn^{\b]} \bar{\f} + 2 \bar{G}_{\s\r} h^{\r[\a} \bn^{\b]} \bar{\f} \nonumber \\
		& \, \qquad + 2 \bar{G}_{\s}^{[\a} h^{\b]\k} \bn_{\k} \bar{\f} - 2 \bar{G}^{[\a}_{\s} \bn^{\b]} \d \f \Big] + \left( \pd_{\f} \bar{\hat{G}}_5 \d \f + \pd_{X} \bar{\hat{G}}_5 \d X + \half \bar{\hat{G}}_5 h \right) \bar{i}^{\a\b}_{\s (G_5)}.
\end{align}
The terms $\d(\nb_{\a\b\g} \f)$, $\d \D^{\l}_{\m\n}$ and $\d G_{\m\n}$ should be replaced with expressions (\ref{d3flinfin}), (\ref{dChdifflin}) and (\ref{dGmnlin}).
\begin{align}
	\hat{R}^{\a\b}_{\s} = & \, 2 \pd_{\f} \bar{\hat{G}}_5 \Big[ \d^{[\a}_{\s} \Big( \hspace{-0.3em} - \hspace{-0.2em} h^{\b]\r} \bn_{\r} \bar{\f} \, \bar{\Box} \bar{\f} + \bn^{\b]} \d \f \, \bar{\Box} \f - \bn^{\b]} \bar{\f} \, h^{\k\l} \bn_{\k\l} \bar{\f} + \bn^{\b]} \bar{\f} (\bar{\Box} \d \f - \bar{g}^{\k\l} \d \D^{\r}_{\k\l} \bn_{\r} \bar{\f})  \nonumber \\
		& \, + h^{\b]\k} \bn_{\k\r} \bar{\f} \, \bn^{\r} \bar{\f} + \bn{^{\b]}}_{\r} \bar{\f} \, h^{\r\l} \bn_{\l} \bar{\f} - \bn^{\b]\r} \d \f \bn_{\r} \bar{\f} + \bar{g}^{\b]\k} \d \D^{\l}_{\k\r} \bn_{\l} \bar{\f} \bn^{\r} \bar{\f} - \bn^{\b]\r} \bar{\f} \, \bn_{\r} \d \f \Big) \nonumber \\
		& \, + \bn^{[\a} \d \f \bn{^{\b]}}_{\s} \bar{\f} - \bn_{\r} \bar{\f} h^{\r[\a} \bn{^{\b]}}_{\s} \bar{\f} - \bn^{[\a} \bar{\f} h^{\b]\k} \bn_{\k\s} \bar{\f} + \bn^{[\a} \bar{\f} \bn{^{\b]}}_{\s} \d \f  \nonumber \\
		& \,  - \bn^{[\a} \bar{\f} \, \bar{g}^{\b]\k} \d \D^{\r}_{\k\s} \bn_{\r} \bar{\f} \Big] + \left( \pd_{\f\f} \bar{\hat{G}}_5 \d \f + \pd_{X\f} \bar{\hat{G}}_5 \d X + \half \pd_{\f} \bar{\hat{G}}_5 \, h \right) \bar{i}^{\a\b}_{\s (\pd_{\f} G_5)},
\end{align}
\begin{align}
	\hat{S}^{\a\b}_{\s} = & \, \pd_{X} \bar{\hat{G}}_5 \left( T^{\a\b}_{\s} + U^{\a\b}_{\s} \right) + \left( \pd_{X\f} \bar{\hat{G}}_5 \d \f + \pd_{XX} \bar{\hat{G}}_5 \d X + \half \pd_{X} \bar{\hat{G}}_5 \, h \right) \bar{i}^{\a\b}_{\s (\pd_X G_5)},
\end{align}
\begin{align}
T^{\a\b}_{\s} =& \, \d^{[\a}_{\s} \Big(  
+ 2\, {h}^{\b] \l} {\bn}^{\r}{\bar{\f}}\,  {\bn}_{\r \l}{\bar{\f}}\,  \bar{\Box} \, \bar{\f}\,   
+ 2\, {\bn}_{\r}{\bar{\f}}\,  {\bn}^{\b]}\,_{\l}{\bar{\f}}\,  \bar{\Box} \, \bar{\f}\,  {h}^{\r \l} 
- 2\, {\bn}^{\b]}\,_{\r}{\d{\f}\, }\, {\bn}^{\r}{\bar{\f}}\,    \bar{\Box} \, \bar{\f}\, \nonumber \\
& 
+ 2\, \bar{g}^{\b] \k} \d{{\D}^{\r}_{\k \l}}\,  {\bn}_{\r}{\bar{\f}}\,  {\bn}^{\l}{\bar{\f}}\,  \bar{\Box} \, \bar{\f}\,  
- 2\, {\bn}^{\b]}\,_{\r}{\bar{\f}}\, {\bn}^{\r}{\d{\f}\, }\,    \bar{\Box} \, \bar{\f}\,  
+ 2\,  {\bn}^{\b]}\,_{\r}{\bar{\f}}\, {\bn}^{\r}{\bar{\f}}\,   {\bn}_{\l \k}{\bar{\f}}\,  {h}^{\l \k} 
\nonumber \\ &
- 2\, {\bn}^{\b]}\,_{\r}{\bar{\f}}\,  {\bn}^{\r}{\bar{\f}}\,  \bar{\Box}{\d{\f}}\,   
+ 2\, {\bn}^{\b]}\,_{\k}{\bar{\f}}\,  \bar{g}^{\l \t} \d{{\D}^{\r}_{\l \t}}\,  {\bn}_{\r}{\bar{\f}}\,  {\bn}^{\k}{\bar{\f}}\,  
- 2\, {h}^{\b] \k}  {\bn}^{\r}{\bar{\f}}\,  {\bn}_{\r}\,^{\l}{\bar{\f}}\,  {\bn}_{\l \k}{\bar{\f}}\,  
\nonumber \\ &
- 2\,  {\bn}^{\b]}\,_{\l}{\bar{\f}}\, {\bn}^{\r}{\bar{\f}}\,  {\bn}_{\r \k}{\bar{\f}}\,  {h}^{\l \k} 
+ 2\, {\bn}^{\b] \l}{\d{\f}\, }\, {\bn}^{\r}{\bar{\f}}\,    {\bn}_{\r \l}{\bar{\f}}\,  
- 2\, \bar{g}^{\b]\t} \d{{\D}^{\r}_{\t \l}}\,  {\bn}_{\r}{\bar{\f}}\,  {\bn}^{\k}{\bar{\f}}\,  {\bn}{^{\l}}_{\k}{\bar{\f}}\,  
\nonumber \\ &
+ 2\, {\bn}^{\b]\l}{\bar{\f}}\,  {\bn}^{\r}{\bar{\f}}\,  {\bn}_{\l\r}{\d{\bar{\f}}\, }\,   
- 2\, {\bn}^{\b] \k}{\bar{\f}}\,  \d{{\D}^{\r}_{\l \k}}\,  {\bn}_{\r}{\bar{\f}}\,  {\bn}^{\l}{\bar{\f}}\,  
- 2\, {\bn}_{\r}{\bar{\f}}\,  {\bn}^{\b] \l}{\bar{\f}}\,  {\bn}_{\l \k}{\bar{\f}}\,  {h}^{\r \k} 
\nonumber \\ &
+ 2\,  {\bn}^{\b \l}{\bar{\f}}\, {\bn}^{\r}{\d{\f}\, }\,   {\bn}_{\r \l}{\bar{\f}}\,  
- {h}^{\b] \r}  {\bn}_{\r}{\bar{\f}}\,  (\bar{\Box} \, \bar{\f})^2 
+ {\bn}^{\b]}{\d{\f}\, }\,  (\bar{\Box} \, \bar{\f})^2  
- 2\, {\bn}^{\b]}{\bar{\f}}\,  \bar{\Box} \, \bar{\f}\,  {\bn}_{\r \l}{\bar{\f}}\,  {h}^{\r \l}
\nonumber \\ &
+ 2\, {\bn}^{\b]}{\bar{\f}}\,  \bar{\Box}{\d{\f}\, }\,  \bar{\Box} \, \bar{\f}\,  
- 2\, {\bn}^{\b]}{\bar{\f}}\, \bar{g}^{\k\l} \d{{\D}^{\r}_{\k\l}}\,  {\bn}_{\r}{\bar{\f}}\,    \bar{\Box} \, \bar{\f}\,  
+ {h}^{\b] \r}  {\bn}_{\r}{\bar{\f}}\,  {\bn}^{\l \k}{\bar{\f}}\,  {\bn}_{\l \k}{\bar{\f}}\,  
\nonumber \\ &
- {\bn}^{\b]}{\d{\f}\, }\,  {\bn}^{\r \l}{\bar{\f}}\,  {\bn}_{\r \l}{\bar{\f}}\,  
- 2\, {\bn}^{\b]}{\bar{\f}}\,  {\bn}^{\r \l}{\d{\f}\, }\,  {\bn}_{\r \l}{\bar{\f}}\,  
+ 2\, {\bn}^{\b]}{\bar{\f}}\, \d{{\D}^{\r}_{\l \k}}\, {\bn}_{\r}{\bar{\f}}\, {\bn}^{\l \k}{\bar{\f}}\,
\nonumber \\ &
+ 2\, {\bn}^{\b]}{\bar{\f}}\,  {\bn}^{\r}\,_{\l}{\bar{\f}}\,  {\bn}_{\r \k}{\bar{\f}}\,  {h}^{\l \k} \Big),
\end{align}
\begin{align}
U^{\a\b}_{\s} =& \, 
- 2\, {\bn}_{\r}{\bar{\f}}\, {h}^{\r [\a}  {\bn}{^{\b]}}_{\s}{\bar{\f}}\,  \bar{\Box} \, \bar{\f}\,  
+ 2\, {\bn}^{[\a}{\d{\f}\, }\,  {\bn}{^{\b]}}_{\s}{\bar{\f}}\,  \bar{\Box} \, \bar{\f}\,  
- 2\, {\bn}^{[\a}{\bar{\f}}\, {h}^{\b] \r}  {\bn}_{\s \r}{\bar{\f}}\,  \bar{\Box} \, \bar{\f}\, \nonumber \\  
& + 2\, {\bn}^{[\a}{\bar{\f}}\,  {\bn}^{\b]}\,_{\s}{\d{\f}\, }\,  \bar{\Box} \, \bar{\f}\,  
+ 2\, \d{{\D}^{\r}_{\s\l}}\, \bar{g}^{\l[\a}  {\bn}^{\b]}{\bar{\f}}\, {\bn}_{\r}{\bar{\f}}\,  \bar{\Box} \, \bar{\f}\,  
- 2\, {\bn}^{[\a}{\bar{\f}}\,  {\bn}{^{\b]}}_{\s}{\bar{\f}}\,  {\bn}_{\r \l}{\bar{\f}}\,  {h}^{\r \l} \nonumber \\ 
& + 2\, {\bn}^{[\a}{\bar{\f}}\, {\bn}_{\s}\,^{\b]}{\bar{\f}}\,    \bar{\Box}{\d{\f}\, }\,  
- 2\, \d{{\D}^{\r}_{\k\l}}\, \bar{g}^{\k\l}  {\bn}_{\r}{\bar{\f}}\,  {\bn}^{[\a}{\bar{\f}}\, {\bn}_{\s}\,^{\b]}{\bar{\f}}\,  
+ 2\, {\bn}_{\r}{\bar{\f}}\,  {\bn}_{\s}\,^{\l}{\bar{\f}}\,  {h}^{\r [\a}   {\bn}^{\b]}\,_{\l}{\bar{\f}}\, \nonumber \\
& - 2\, {\bn}^{[\a}{\d{\f}\, }\, {\bn}^{\b]}\,_{\r}{\bar{\f}}\, {\bn}_{\s}\,^{\r}{\bar{\f}}\,    
+ 2\, {\bn}^{[\a}{\bar{\f}}\, {h}^{\b] \l}   {\bn}_{\s}\,^{\r}{\bar{\f}}\,  {\bn}_{\r \l}{\bar{\f}}\,  
+ 2\, {\bn}^{[\a}{\bar{\f}}\,  {\bn}^{\b]}\,_{\l}{\bar{\f}}\,  {\bn}_{\s \r}{\bar{\f}}\,   {h}^{\r \l} \nonumber \\
& - 2\, {\bn}^{[\a}{\bar{\f}}\,  {\bn}^{\b] \r}{\d{\f}\, }\,  {\bn}_{\s \r}{\bar{\f}}\,  
- 2\, \d{{\D}^{\r}_{\k \l}}\, \bar{g}^{\k[\a} {\bn}^{\b]}{\bar{\f}}\, {\bn}_{\r}{\bar{\f}}\, {\bn}{_{\s}}^{\l}{\bar{\f}}\,  
- 2\, {\bn}^{[\a}{\bar{\f}}\, {\bn}^{\b]}\,_{\r}{\bar{\f}}\, {\bn}^{\r}\,_{\s}{\d{\f}\, }\, \nonumber \\   
& + 2\, \d{{\D}^{\r}_{\s\l}}\,  {\bn}_{\r}{\bar{\f}}\,  {\bn}^{[\a}{\bar{\f}}\,  {\bn}^{\b]\l}{\bar{\f}}
+ 2\, {\bn}^{\r}{\d{\f}\, }\,  {\bn}_{\s}\,^{[\a}{\bar{\f}}\,  {\bn}^{\b]}\,_{\r}{\bar{\f}}\,  
- 2\, {\bn}^{\r}{\bar{\f}}\,  {\bn}_{\s \l}{\bar{\f}}\, {h}^{\l[\a}   {\bn}^{\b]}\,_{\r}{\bar{\f}}\, \nonumber \\
& + 2\, {\bn}^{\r}{\bar{\f}}\,  {\bn}_{\s}\,^{[\a}{\d{\f}\, }\,  {\bn}^{\b]}\,_{\r}{\bar{\f}}\,
- 2\, \d{{\D}^{\r}\,_{\s\k}}\, \bar{g}^{\k[\a} {\bn}^{\b]}\,_{\l}{\bar{\f}}\,   {\bn}_{\r}{\bar{\f}}\,  {\bn}^{\l}{\bar{\f}}\,  
- 2\, {\bn}^{\r}{\bar{\f}}\,  {\bn}_{\s}\,^{[\a}{\bar{\f}}\, {h}^{\b] \l}  {\bn}_{\r \l}{\bar{\f}}\, \nonumber \\
& - 2\, {\bn}_{\r}{\bar{\f}}\,  {\bn}_{\s}\,^{[\a}{\bar{\f}}\,  {\bn}^{\b]}\,_{\l}{\bar{\f}}\,  {h}^{\r \l} 
+ 2\, {\bn}^{\r}{\bar{\f}}\, {\bn}_{\s}\,^{[\a}{\bar{\f}}\,  {\bn}^{\b]}\,_{\r}{\d{\f}\, }\,   
+ 2\, \d{{\D}^{\r}_{\k \l}}\, \bar{g}^{\k[\a} {\bn}{^{\b]}}_{\s}{\bar{\f}}\,    {\bn}_{\r}{\bar{\f}}\,  {\bn}^{\l}{\bar{\f}}.
\end{align}
For $\bar{\f} = \mbox{const.}$, the preceding results simplify into
\begin{align}
\d \hat{i}^{\a\b}_{(5)(rest)} =& \, \bar{\hat{G}}_5 \Big[ 2 \left( \d^{[\a}_{\s} \bn{^{\b]\l}}_{\l} \d \f - \d^{[\a}_{\s} \bn{_{\l}}^{\b]\l} \d \f + \bn{^{[\a\b]}}_{\s} \d \f - G^{[\a}_{\s} \bn^{\b]} \d \f \right) \x^{\s} \nonumber \\
	& \, \quad + \left( \d^{[\a}_{\s} \bn^{\b]\l} \d \f - \bar{g}^{\l[\a} \bn{^{\b]}}_{\s} \d \f \right) \bn_{\l} \x^{\s} \Big].
\end{align}

\section{Brans-Dicke theory}

Considering the Brans-Dicke Lagrangian
\begin{equation}
	L = \sqrt{-g} \left( \half \f R + \frac{\omega}{\f} X - U(\f) \right),
\end{equation}
all of the results considerably simplify. The Brans-Dicke theory is a special case of a general Horndeski theory with Lagrangians $\Lh_2$ and $\Lh_4$ with functions $K$ and $G_4$ given as follows:
\begin{align}
	K(\f, X) &= \frac{\omega}{\f} X - U(\f), & G_4(\f) &= \half \f.
\end{align}

The non-vanishing superpotential associated with the Lagrangian $\Lh_4$ is obtained after setting $\pd_{\f} G_4 = \half$ and $\pd_X G_4 = 0$ in (\ref{iab4}) and (\ref{iab4rest}) with the result:
\begin{equation}
	\hat{i}^{\a\b}_{(BD)} = \half \varphi \, \hat{i}^{\a\b}_{(EH)} - 2 \sqrt{-g} \, \d^{[\a}_{\s} \nb^{\b]} \f \, \x^{\s}.
\end{equation}

The superpotential associated with the background reads
\begin{equation}
	\bar{\hat{i}}^{\a\b}_{(BD)} = \half \bar{\f} \, \bar{\hat{i}}^{\a\b}_{(EH)} - 2 \sqrt{-\bar{g}} \, \d^{[\a}_{\s} \bn^{\b]} \bar{\f} \, \x^{\s}.
\end{equation}

Finally, the linearization of the superpotential becomes:
\begin{equation}
	\d \hat{i}^{\a\b}_{(BD)} = \half \d \f \, \bar{\hat{i}}^{\a\b}_{(EH)} + \half \bar{\f} \, \d \hat{i}^{\a\b}_{(EH)} - 2 \sqrt{-\bar{g}} \, \d^{[\a}_{\s} \left( \half h \bn^{\b]} \bar{\f} - h^{\b]\r} \bn_{\r} \bar{\f} + \bn^{\b]} \d \f  \right) \x^{\s}.
\end{equation}

\begin{acknowledgments}
J.B. acknowledges the support from the Czech Science Foundation, GA\v{C}R Grant No. 14-37086G (Albert Einstein Centre); J.S. was supported by the Grant Agency of the Czech Technical University in Prague, Grant No. SGS13/217/OHK4/3T/14. We also thank for the hospitality of Albert Einstein Institute in Golm where we enjoyed brief but useful collaboration.
\end{acknowledgments}

\appendix

\section{Conserved current coefficients for the Einstein-Hilbert Lagrangian}

Considering the Einstein-Hilbert Lagrangian density, $\L_{(EH)} = \hat{R} = \sqrt{-g} \, R$, we can calculate the coefficients using the formulas (\ref{gcurn})--(\ref{gcuru}) with $\Lh_{(EH)}$. Concerning $\hat{u}^{\a}_{\s}$ we have to add the term $\L \d^{\a}_{\s}$ to (\ref{gcuru}) which we explicitly excluded in (\ref{splitcoef}) to be able to split the coefficients into scalar and tensor parts. The conserved current coefficients involve partial derivatives of the Riemann tensor with respect to the metric tensor and its (background) covariant derivatives. These were already considered in the currents associated with the Horndeski Lagrangians $\Lh_4$ and $\Lh_5$. For the derivative of the scalar curvature with respect to the second derivative of the metric, see (\ref{rd2}), the derivative of the Ricci tensor w.r.t. the first derivative of the metric tensor can be found by contracting (\ref{ricd1}):
\begin{equation}
	\frac{\pd R}{\pd g_{\m\n|\a}} = \D^{\a}_{\r\k} \left( g^{\r\k} g^{\m\n} - g^{\r(\m} g^{\n)\k} \right) + 2 \D^{(\m}_{\r\k} \left( g^{\n)\k} g^{\a\r} - g^{\n)\a} g^{\r\k} \right).
\end{equation}

For the coefficients $\hat{n}^{\a\t\b}_{\s(EH)}$, $\hat{m}^{\a\t}_{\s(EH)}$ and $\hat{u}^{\a}_{\s(EH)}$ we get
\begin{align}
	\hat{n}^{\a\t\b}_{\s(EH)} =& \, \d^{(\b}_{\s} \hat{g}^{\t)\a} - \d^{\a}_{\s} \hat{g}^{\b\t}, \\
	\hat{m}^{\a\t}_{\s(EH)} =& \, \D^{\t}_{\r\k} \hat{g}^{\r\k} \d^{\a}_{\s} + \D^{\r}_{\r\s} \hat{g}^{\a\t} - 2 \D^{\a}_{\s\r} \hat{g}^{\r\t}, \\
	\hat{u}^{\a}_{\s(EH)} =& \, \hat{R} \d^{\a}_{\s} - 2 \hat{G}{^{\a}}_{\s} + 2 \hat{g}^{\a\r} \bn_{\r} \D^{\k}_{\k\s} - \hat{g}^{\r\k} \bn_{\r} \D^{\a}_{\k\s} - \hat{g}^{\a\r} \bn_{\k} \D^{\k}_{\r\s} \nonumber \\
		& \, +\hat{g}^{\a\r} \D^{\k}_{\r\l} \D^{\l}_{\k\s} - \hat{g}^{\a\r} \D^{\k}_{\r\s} \D^{\l}_{\k\l} - \D^{\a}_{\r\k} \D^{\r}_{\s\l} \hat{g}^{\k\l} + \D^{\a}_{\s\r} \D^{\r}_{\k\l} \hat{g}^{\k\l} - \frac{3}{2} \hat{g}^{\a\r} \bar{R}_{\r\s}.
\end{align}

The general formula for the superpotential (\ref{superpot}) requires the antisymmetrized current coefficients,
\begin{align}
	\hat{n}^{[\a\b]\l}_{\s(EH)} &= - \frac{3}{2} \d^{[\a}_{\s} \hat{g}^{\b]\l}, \\
	\hat{m}^{[\a\b]}_{\s(EH)} &= \d^{[\a}_{\s} \D^{\b]}_{\r\k} \hat{g}^{\r\k} - 2 \D^{[\a}_{\s\r} \hat{g}^{\b]\r},
\end{align}
and, regarding the relation $\bn_{\a} \hat{T}^{\cdots}_{\cdots} = \D^{\l}_{\l\a} \hat{T}^{\cdots}_{\cdots} + \sqrt{-g} \, \bn_{\a} T^{\cdots}_{\cdots}$, we obtain the resulting superpotential for the Einstein-Hilbert Lagrangian in the form
\begin{align}
	\hat{i}^{\a\b}_{(EH)} &= 2 \D^{[\a}_{\s\l} \hat{g}^{\b]\l} \x^{\s} + 2 \d^{[\a}_{\s} \hat{g}^{\b]\l} \bn_{\l} \x^{\s}.
\end{align}
Adding a divergence of a vector density $\pd_{\a} \hat{d}^{\a}$ to a Lagrangian does not change the equations of motion, but it changes the conserved current coefficient $\hat{m}^{\a\b}_{\s}$ and corresponding superpotential $\hat{i}^{\a\b}$ as follows:
\begin{align}
	\hat{m}^{\a\b}_{\s} &\rightarrow \hat{m}^{\a\b}_{\s} + 2 \d^{[\a}_{\s} \hat{d}^{\b]}, & \hat{i}^{\a\b} \rightarrow \hat{i}^{\a\b} - 2 \d^{[\a}_{\s} \hat{d}^{\b]} \xi^{\s},
\end{align}
see Ref. \citen{petrovmain}, Eq. (84). Considering the vector density $\hat{d}^{\m} = \hat{k}^{\m} = \hat{g}^{\m\r} \D^{\k}_{\r\k} - \hat{g}^{\r\k} \D^{\m}_{\r\k}$ as in Ref. \citen{KBL}, the following conserved current and superpotential modification, denoted as $\hat{m}^{\a\b}_{\s(k)}$ and $\hat{i}^{\a\b}_{(k)}$ arise:
\begin{align}
	\hat{m}^{\a\b}_{\s(k)} &= 2 \d^{[\a}_{\s} \hat{g}^{\b]\r} \D^{\k}_{\r\k} - 2 \d^{[\a}_{\s} \D^{\b]}_{\r\s} \hat{g}^{\r\s}, & \hat{i}^{\a\b}_{(k)} &= - 2 \d^{[\a}_{\s} \hat{g}^{\b]\r} \D^{\k}_{\r\k} \x^{\s} + 2 \d^{[\a}_{\s} \D^{\b]}_{\r\k} \hat{g}^{\r\k} \x^{\s}.
\end{align}
If the divergence of the vector density $\hat{k}^{\m}$ is added to the Einstein-Hilbert Lagrangian, the KBL superpotential \cite{KBL} is recovered:
\begin{equation}
	- 2 \, \hat{i}^{\a\b}_{(KBL)} = \hat{i}^{\a\b}_{(EH)} + \hat{i}^{\a\b}_{(k)}.
\end{equation}
If we wish to recover the KBL superpotential from a general Horndeski theory this divergence has to be added to the general Horndeski Lagrangian.

\nocite{*}

\end{document}